%% file: main.tex
\pdfoutput=1
\documentclass[11pt]{article}
\usepackage{style}

\title{Lower Bounds Above $\sqrt{3}$ for Online Makespan Minimization on Four Identical Machines}
\author{Maximilian Janke\thanks{Google, Mountain View, CA, USA. Email: \texttt{maxjanke@google.com}.}}
\date{}

\begin{document}
\maketitle

\begin{abstract}
For deterministic online makespan minimization, $m = 4$ identical machines is the smallest case whose optimal competitive ratio is open: it is known only to lie between $\sqrt{3} \approx 1.73205$ and $\tfrac{26}{15} \approx 1.73333$. We prove a lower bound of $\tfrac{14383}{8304} \approx 1.73206$ using a single branching layer, and refine the construction with a second, mirrored branching layer to obtain $\tfrac{14350}{8283} \approx 1.73246$, closing about a third of the gap.

Our constructions are the first branching adversaries for $m = 4$ machines, and the first explicit lower bounds for online makespan minimization in which the adversary presents jobs in non-monotone order of size.
\end{abstract}

\section{Introduction}

Makespan minimization on parallel, identical machines is one of the foundational problems in scheduling theory and competitive analysis~\cite{Graham1966}: jobs arrive one by one, and each job must be assigned immediately to one of $m$ machines so as to keep the maximum machine load within a factor~$c$ of the offline optimum. Graham~\cite{Graham1966} established the optimal deterministic competitive ratio $\tfrac{3}{2}$ for $m = 2$ machines in 1966, and Faigle, Kern, and Tur\'an~\cite{Faigle1989} established $\tfrac{5}{3}$ for $m = 3$ in 1989. On $m = 4$ machines, by contrast, the best known upper bound of $\tfrac{26}{15}$ has stood since 1994~\cite{Chen1994} and the lower bound of $\sqrt{3}$ since 2001~\cite{Rudin2001,Rudin2003}. Like every previous lower bound for $m = 4$ machines (and every explicit layered construction for general~$m$), the $\sqrt{3}$ bound uses a \emph{linear} adversary: the job sequence is fixed in advance, and the online algorithm's choices affect only when the sequence stops. In this paper, we introduce \emph{branching} adversaries for $m = 4$ machines that choose the sizes of future jobs depending on where the online algorithm placed earlier ones, obtaining the first lower bounds strictly above~$\sqrt{3}$.

\subsection{Related Work}\label{sec:related}

Graham~\cite{Graham1966} showed that his List Scheduling (Greedy) strategy is $(2 - \tfrac{1}{m})$-competitive. Galambos and Woeginger~\cite{Galambos1993} beat this bound for every $m \ge 4$ by preventing schedules from becoming flat, though their improvement vanishes as $m$ grows. Bartal, Fiat, Karloff, and Vohra~\cite{Bartal1995} gave the first algorithm whose competitive ratio stays bounded away from~$2$ for all~$m$, achieving $2 - \tfrac{1}{70} \approx 1.986$ by keeping a constant fraction of the machines lightly loaded. Subsequent improvements to $1.945$ by Karger, Phillips, and Torng~\cite{Karger1996} and to $1.923$ by Albers~\cite{Albers1999} culminated in the $1.9201$-competitive algorithm of Fleischer and Wahl~\cite{Fleischer2000}, which remains the best known deterministic upper bound for large~$m$.

On the lower bound side, Faigle, Kern, and Tur\'an~\cite{Faigle1989} proved a bound of $1 + \tfrac{1}{\sqrt{2}} \approx 1.7071$ for $m \ge 4$. For large~$m$, Bartal, Karloff, and Rabani~\cite{Bartal1994} established $1.837$, Albers~\cite{Albers1999} obtained $1.852$, Gormley, Reingold, Torng, and Westbrook~\cite{Gormley2000} reached $1.853$ via a computer search over request-answer game graphs, and Rudin~\cite{Rudin2001} proved the current best bound of~$1.88$. Apart from the computer-generated game graphs of~\cite{Gormley2000}, all of these constructions use layered linear adversaries (\Cref{sec:linear}); as noted by Rudin~\cite{Rudin2001}, a natural family of layered linear adversaries cannot exceed $\tfrac{17}{9} \approx 1.889$. For randomized algorithms against an oblivious adversary, the best known upper and lower bounds are $1.916$~\cite{Albers2002} and $\tfrac{e}{e-1} \approx 1.582$~\cite{Chen1994rand,Sgall1997}. In the random-order model, Albers and Janke~\cite{Albers2021} gave a deterministic algorithm that, for every job set, is $(1.8478+\varepsilon_m)$-competitive on a $(1-\varepsilon_m)$-fraction of its arrival orders, where $\varepsilon_m \to 0$ as $m \to \infty$; adversarial instances forcing ratios near $1.88$ are thus rare and sensitive to the arrival order.

For $m = 4$ machines specifically, Chen, van Vliet, and Woeginger~\cite{Chen1994} improved the upper bound of Galambos and Woeginger~\cite{Galambos1993} to $\tfrac{26}{15}$ and proved a lower bound of $1.7310$ above the $1 + \tfrac{1}{\sqrt{2}}$ bound of~\cite{Faigle1989}, before Rudin~\cite{Rudin2001} and Rudin and Chandrasekaran~\cite{Rudin2003} established $\sqrt{3}$. For any fixed~$m$, L\"ubbecke, Maurer, Megow, and Wiese~\cite{Lubbecke2016} and Chen, Ye, and Zhang~\cite{Chen2015} gave $(1 + \varepsilon)$-approximation schemes for the optimal competitive ratio via game-tree exploration and dynamic programming, respectively.

\paragraph{Online bin packing.}
In online bin packing, items of size in $(0, 1]$ arrive online to be packed into unit-capacity bins. Classical linear adversaries yielded asymptotic lower bounds of~$1.5$~\cite{Yao1980}, $1.536$~\cite{Liang1980}, $1.5401$~\cite{vanVliet1992}, and $\tfrac{248}{161} \approx 1.5403$~\cite{Balogh2012}, before Balogh, B\'ek\'esi, D\'osa, Epstein, and Levin~\cite{Balogh2021} introduced the first branching lower bound for classic online bin packing to obtain $1.54278$. The best known upper bound is $1.5783$~\cite{Balogh2018}, whereas the absolute competitive ratio is settled at $\tfrac{5}{3}$~\cite{Balogh2019}.

\paragraph{Semi-online and incremental models.}
In semi-online variants of makespan minimization, the algorithm receives partial advance information about the input. When the total processing time is given upfront, a lower bound of~$1.585$ for large~$m$~\cite{Albers2012semi} matches the $1.585$-competitive algorithm for all~$m$~\cite{Kellerer2015}; when jobs arrive in nonincreasing order of size, Largest Processing Time first (LPT) is $(\tfrac{4}{3} - \tfrac{1}{3m})$-competitive~\cite{Graham1969}, with tighter online bounds for $m = 2, 3$~\cite{Seiden2000}. When the optimum makespan is known in advance (\emph{online bin stretching}), Azar and Regev~\cite{Azar2001} proved a lower bound of~$\tfrac{4}{3}$ for all $m \ge 2$, while the best upper bounds are~$1.5$ in general~\cite{Bohm2017twophase} and $\tfrac{139}{93} < 1.495$ for large~$m$~\cite{Bohm2024}. For small~$m$, computer searches over the game tree have produced branching adversaries with non-monotone probes reaching $\tfrac{56}{41} \approx 1.366$ for $m = 3$~\cite{Bohm2017three,Gabay2017,Bohm2022}, $\tfrac{19}{14} \approx 1.357$ for $m = 4, 5$~\cite{Gabay2017,Bohm2017three}, and $\tfrac{15}{11} \approx 1.364$ for $m = 6, 7, 8$~\cite{Bohm2022,Lhomme2022}. At the extreme where the entire input sequence is known in advance, \emph{incremental} algorithms build their solution step by step while remaining $c$-competitive on every prefix~\cite{Mettu2003,Chrobak2008,Lin2010,Bernstein2022}; as we note in \Cref{sec:linear}, for makespan minimization their power coincides with that of linear adversaries.

\subsection{Our Contribution}\label{sec:contribution}

We prove two lower bounds strictly above $\sqrt{3}$ for deterministic online makespan minimization on $m = 4$ identical machines. To make the mechanism transparent, we first present a simpler strategy consisting of a five-layer linear prefix, a single branching layer with four cases, and a linear endgame.

\begin{theorem}\label{thm:easy}
No deterministic online algorithm for makespan minimization on $m = 4$ identical machines achieves a competitive ratio smaller than
\[
  c_0 \;=\; \frac{14383}{8304} \;\approx\; 1.73205684 \;>\; \sqrt{3} \;\approx\; 1.73205081.
\]
\end{theorem}

Retaining the exact machine loads from the five-layer prefix (rather than cutting off excess load) and adding a second, mirrored branching layer yields our main result.

\begin{theorem}\label{thm:opt}
No deterministic online algorithm for makespan minimization on $m = 4$ identical machines achieves a competitive ratio smaller than
\[
  c \;=\; \frac{14350}{8283} \;\approx\; 1.73246408 \;>\; \sqrt{3} \;\approx\; 1.73205081.
\]
\end{theorem}

In a layered linear adversary (\Cref{sec:linear}), jobs arrive in layers of four that lift the minimum online load from one predetermined level to the next. In early layers there are few jobs, so offline schedules cannot pack them tightly; in later layers the offline schedule has more flexibility, so the last job of a layer must be enlarged to prevent the online algorithm from placing two jobs of the layer on one machine (\Cref{ex:linear}). The machine receiving this larger job becomes heavier than the other three, and although the offline schedule must pack the extra load, a linear adversary cannot exploit this extra load in subsequent layers. Our branching adversaries exploit this load gap: after a linear prefix creates a \emph{heavy machine}, the adversary presents decreasing \emph{probes} until the online algorithm places one on the heavy machine, and then chooses the remaining jobs of the layer to equalize (or nearly equalize) the machine loads.

In previous lower bounds for online makespan minimization, jobs arrive in nondecreasing order of size~\cite{Albers2021}, matching the intuition that presenting a smaller job after a larger one only helps the online algorithm. In our constructions, job sizes still grow from layer to layer, but the probes within the first branching layer decrease slightly: if the online algorithm delays using the heavy machine, a slightly smaller probe suffices to bring that machine to the target load while tightening the offline optimum on the preceding light-machine placements (restricting the strategy of \Cref{thm:easy} to nondecreasing job sizes reaches at most $\approx 1.7319 < \sqrt{3}$). In the second branching layer of \Cref{thm:opt}, where the distinguished machine is lighter than the other three, the probes increase again; lengthening the linear prefix or using larger integer weights in this two-layer branching scheme yields only a marginal third-decimal-place gain at the cost of a larger case analysis.

\Cref{sec:defs,sec:linear} formalize the adversarial game and review linear adversaries; \Cref{sec:easy,sec:opt} prove \Cref{thm:easy,thm:opt}, with step-by-step verification tables in \Cref{sec:appendix_tables}.

\subsection{Problem Definition}\label{sec:defs}

In online makespan minimization on $m = 4$ identical machines, a finite job sequence $\CJ = J_1, \dots, J_n$ arrives one job at a time. Each job $J_t$ is specified by its positive \emph{processing time} (or \emph{size}) $p_t > 0$ and must be assigned irrevocably and non-preemptively to one of the four machines before $J_{t+1}$ is revealed. The \emph{load} of a machine is the sum of the sizes of the jobs assigned to it, and the \emph{makespan} of a schedule is the maximum load over all four machines. We write $\ALG(\CJ)$ for the makespan produced by a deterministic online algorithm $\ALG$ on $\CJ$, and $\OPT(\CJ)$ for the minimum makespan of an offline schedule of $\CJ$. When $\CJ$ is clear from context we write $\ALG$ and $\OPT$ for $\ALG(\CJ)$ and $\OPT(\CJ)$. An online algorithm $\ALG$ is \emph{$c$-competitive} if $\ALG(\CJ) \le c \cdot \OPT(\CJ)$ for every finite sequence~$\CJ$.

\paragraph{Lower bounds as a game.}
By machine symmetry, the state of $\ALG$ after any step is determined by its sorted quadruple of machine loads $(\lambda_1, \lambda_2, \lambda_3, \lambda_4)$ with $\lambda_1 \ge \lambda_2 \ge \lambda_3 \ge \lambda_4$, where $\lambda_1$ is the current makespan of $\ALG$. Lower bounds are obtained via a round-by-round game between $\ALG$ and an adversary for a target ratio~$c$. Initially, all four loads are $0$. In round~$t = 1, 2, \dots$, the adversary observes the current loads $(\lambda_1, \lambda_2, \lambda_3, \lambda_4)$ of $\ALG$ and presents a job $J_t$ of size $p_t > 0$, which $\ALG$ assigns to one of the machines. The adversary \emph{wins} and stops as soon as
$\lambda_1 \ge c \cdot \OPT(J_1, \dots, J_t)$.
A strategy that wins against every deterministic online algorithm certifies a lower bound of~$c$. To reduce the number of states to distinguish, our analysis tracks componentwise lower bounds on the machine loads of $\ALG$; equivalently, the adversary may \emph{reduce} (or cut off) the tracked load of any machine to a smaller nonnegative value whenever convenient, as in \Cref{fig:linear_adversary}(a). Since every machine's actual load is at least its reduced load while $\OPT$ is unchanged, any strategy that wins with respect to the reduced loads also wins with respect to the actual loads.

\subsection{Linear Adversaries}\label{sec:linear}
The classic lower bounds for $m = 4$ machines use \emph{layered linear adversaries}, illustrated in \Cref{fig:linear_adversary}, which consist of $k$ \emph{layers} $i = 0, \dots, k-1$ of four jobs each ($J_{4i+1}, \dots, J_{4i+4}$), followed by a single final job $J_{4k+1}$. At the start of layer~$i$, all four machines have the same reduced load~$l_i$, with $l_0 = 0$.

\input{tikz/fig_linear_adversary}
The adversary presents the four jobs $J_{4i+1}, \dots, J_{4i+4}$ of layer~$i$ in nondecreasing order of size $p := p_{4i+1} \le p_{4i+2} \le p_{4i+3} \le p_{4i+4}$, so that placing two jobs of layer~$i$ on one machine is an immediate loss:
\begin{equation}\label{eq:linear_layer}
  l_i + p + p_{4i+j} \;\ge\; c \cdot \OPT(J_1, \dots, J_{4i+j}) \qquad \text{for } j = 2, 3, 4.
\end{equation}
Since appending a job cannot decrease~$\OPT$, we have $\OPT(J_1, \dots, J_{4i+2}) \le \OPT(J_1, \dots, J_{4i+3})$, so whenever $p_{4i+1} = p_{4i+2} = p_{4i+3}$, condition~\eqref{eq:linear_layer} for $j = 3$ subsumes the case $j = 2$. Any machine that already holds a job of layer~$i$ has load at least $l_i + p$, so assigning $J_{4i+j}$ to it raises its load to at least $l_i + p + p_{4i+j}$, and the adversary wins by~\eqref{eq:linear_layer}. Consequently, $\ALG$ must assign each job $J_{4i+j}$ to a machine of load~$l_i$ that has not yet received a job of layer~$i$, and the adversary immediately reduces the resulting load $l_i + p_{4i+j}$ to $l_{i+1} := l_i + p$. More generally, if the four machines have arbitrary loads at the start of layer~$i$ with minimum load $\underline{l}_i = \lambda_4$, then replacing $l_i$ by $\underline{l}_i$ in~\eqref{eq:linear_layer} still forces the four jobs of layer~$i$ onto four distinct machines. Following layer~$k-1$, the adversary presents a final job $J_{4k+1}$ satisfying
\begin{equation}\label{eq:linear_final}
  l_k + p_{4k+1} \;\ge\; c \cdot \OPT(J_1, \dots, J_{4k+1}).
\end{equation}
Since every machine has load at least~$l_k$, the makespan of $\ALG$ reaches at least $l_k + p_{4k+1}$ regardless of where $J_{4k+1}$ is placed, and the adversary wins. We call such an adversary \emph{linear} because, up to machine symmetry and load reductions, $\ALG$ has a unique non-losing move in every round, so the non-terminal game tree is a single path.

\begin{example}\label{ex:linear}
We extend the two-layer adversary of Faigle, Kern, and Tur\'an~\cite{Faigle1989} by one layer. Let
\[
  x = \frac{1+\sqrt{17}}{2} \approx 2.56, \qquad a = 2x + 3 = 4 + \sqrt{17} \approx 8.12, \qquad b = a + 2, \qquad c = \frac{11-\sqrt{17}}{4} \approx 1.7192.
\]
The adversary has $k = 3$ layers and a final job; see \Cref{fig:linear_adversary}.
\begin{itemize}
  \item \emph{Layer~$0$} ($l_0 = 0$): four jobs of size~$1$. Condition~\eqref{eq:linear_layer} holds since $\OPT = 1$ and $0 + 1 + 1 = 2 \ge c$, so $l_1 = 1$.
  \item \emph{Layer~$1$} ($l_1 = 1$): four jobs of size~$x$. Pairing each job of size~$x$ with a job of size~$1$ gives $\OPT \le 1 + x$, and~\eqref{eq:linear_layer} holds with equality: $1 + 2x = c\,(1 + x)$. Thus $l_2 = L := 1 + x$.
  \item \emph{Layer~$2$} ($l_2 = L$): three jobs of size~$a$ followed by one job of size~$b$. For the jobs of size~$a$, the four machine bundles $a + x$, $a + 2 \times 1$, $a + 2 \times 1$, and $3 \times x < a + x$ witness $\OPT \le a + x$, and $L + 2a > c\,(a + x)$. For the job of size~$b$, the bundles $a + x + 2 \times 1$, $a + x + 2 \times 1$, $a + x$, and $b + x$ witness $\OPT \le a + x + 2$, and~\eqref{eq:linear_layer} holds with equality: $L + a + b = c\,(a + x + 2)$. The adversary reduces the load $L + b$ to $l_3 = L + a$.
  \item \emph{Final job}: $p_{13} = 2a$. In the offline schedule, $J_{13}$ runs alone, two jobs of size~$a$ share a second machine, and the remaining jobs fill the other two machines with bundles $a + 2 \times x + 3 \times 1$ and $b + 2 \times x + 1$, each of load $a + 2x + 3 = 2a$. Thus $\OPT = 2a$, and~\eqref{eq:linear_final} holds with equality: $l_3 + 2a = L + 3a = c \cdot 2a$.
\end{itemize}
Enlarging the last job of layer~$2$ is essential: if it also had size~$a$, pairing each job of size~$a$ with $x + 1$ would give $\OPT \le L + a$, and since $L + 2a < c\,(L + a)$, $\ALG$ could place two jobs of size~$a$ on one machine. Increasing the fourth job by~$2$ raises the left-hand side of~\eqref{eq:linear_layer} by~$2$ while raising $\OPT$ by only~$1$, at the cost of the excess $b - a = 2$, which the adversary cuts off and never uses again.
\end{example}

With fewer layers, the same scheme yields $\tfrac{3}{2}$ for $k = 1$ (four jobs of size~$1$ and a final job of size~$2$) and the bound $1 + \tfrac{1}{\sqrt{2}} \approx 1.7071$ of Faigle, Kern, and Tur\'an~\cite{Faigle1989} for $k = 2$ with $x = 1 + \sqrt{2}$ and $p_9 = 2x$. With more layers, Rudin and Chandrasekaran~\cite{Rudin2003} showed that a linear adversary wins for every $c < \sqrt{3}$.

Because a linear adversary fixes its job sequence in advance, knowing the sequence upfront does not help against it: every linear lower bound carries over to the incremental model (\Cref{sec:related}), and conversely the optimal incremental competitive ratio equals the best lower bound achievable by a linear adversary. Thus knowing the job sequence in advance does not help for $m = 2, 3$, and for $m = 4$ the optimal incremental competitive ratio lies in \mbox{$[\sqrt{3}, \tfrac{26}{15}]$}. To our knowledge, for no~$m$ has an incremental algorithm been shown to beat the best online algorithm, and for $m = 4$ we conjecture that the optimal incremental competitive ratio is~$\sqrt{3}$. If so, no linear adversary can beat $\sqrt{3}$, and branching is necessary.

\section{Proof of the Simple Bound (\texorpdfstring{\Cref{thm:easy}}{Theorem 1})}\label{sec:easy}

To prove \Cref{thm:easy}, we specify a winning strategy for the adversary against an arbitrary deterministic online algorithm $\ALG$ over at most $29$ jobs: $k = 7$ layers of four jobs each ($J_1, \dots, J_{28}$), followed by a final job $J_{29}$. The strategy has three phases: \emph{Phase~1} (layers $0$--$4$, Steps~1--20) is a linear prefix in which the last job $J_{20}$ is not reduced, leaving one heavy machine; \emph{Phase~2} (layer~$5$, Steps~21--24) is a branching layer whose job sizes adapt to $\ALG$'s placements to equalize all four loads; and \emph{Phase~3} (layer~$6$ and the final job, Steps~25--29) is a linear endgame. \Cref{tab:easy_strategy} lists all job sizes and the immediate-loss inequalities~\eqref{eq:linear_layer} and~\eqref{eq:linear_final}, \Cref{fig:easy_step20,fig:easy_phase2,fig:easy_endgame} illustrate the three phases, and every upper bound on $\OPT$ is witnessed by an explicit offline schedule in \Cref{tab:easy_opt} (cited as Rows~1--18). Since reducing job sizes cannot increase $\OPT$, one schedule can witness a bound for several cases at once (e.g., Row~9 for Cases~2--4). From now on, we identify each job~$J_t$ with its size~$p_t$ and write, e.g., $J_{20} = 316$.

\input{tables/tab06_easy_strategy}
\input{tikz/fig_simple_step20}

\paragraph{Phase 1: a linear prefix (Steps 1--20).}
The adversary presents the five layers
\[
  (2, 2, 2, 2), \qquad (6, 6, 6, 6), \qquad (20, 20, 20, \mathbf{24}), \qquad (99, 99, 99, 99), \qquad (276, 276, 276, \mathbf{316}).
\]
\Cref{tab:easy_strategy}(a) verifies condition~\eqref{eq:linear_layer} with $c = c_0$ for each layer, forcing the four jobs of each layer onto distinct machines. As in \Cref{ex:linear}, layers~$2$ and~$4$ enlarge their fourth job to maintain~\eqref{eq:linear_layer} for $j = 4$ (in Step~19, three jobs of size~$99$ pack onto one machine of load $297 < 392$, Row~6, leaving only limited slack before Step~20), whereas layer~$3$ is uniform because in Step~15 each job of size~$99$ pairs with one machine of the Row-4 schedule of load at most~$30$ (Row~5). Only $J_{12}$ in layer~$2$ is reduced (from $8 + 24 = 32$ to $l_3 = 28$), so layers~$1$--$4$ start from the common loads $l_1 = 2$, $l_2 = 8$, $l_3 = 28$, and $l_4 = 127$. In layer~$4$, three machines receive $276$ to reach $127 + 276 = 403$, and $J_{20} = 316$ goes to the fourth machine, the \emph{heavy machine}~$H$, whose load $127 + 316 = 443$ is not reduced. Thus, after Step~20:
\begin{itemize}[nosep]
  \item $\ALG$ has reduced loads $(443, 403, 403, 403)$, with $H$ sitting $40$ units above the three \emph{light} machines;
  \item $\OPT(J_1, \dots, J_{20}) \le 415$ via the schedule $(415, 415, 415, 411)$ in Row~7, shown in \Cref{fig:easy_step20}(b), and the $20$ jobs partition into three $552$-blocks, shown in \Cref{fig:easy_step20}(c) and used in Rows~10--18 of \Cref{tab:easy_opt}: $316 + 2 \times 99 + 20 + 2 \times 6 + 3 \times 2$, $276 + 2 \times 99 + 24 + 2 \times 20 + 2 \times 6 + 2$, and $2 \times 276$.
\end{itemize}
Without the reduction in layer~$2$, $\ALG$ would end Phase~1 in $(447, 403, 403, 403)$ or $(443, 407, 403, 403)$ depending on whether $J_{20}$ lands on the machine of $J_{12}$; here the reduction merges both outcomes into $(443, 403, 403, 403)$.

\input{tables/tab01_easy_opt}
\input{tikz/fig_simple_branching}

\paragraph{Phase 2: a branching layer (Steps 21--24).}
To close the $40$-unit gap on~$H$, the adversary presents decreasing \emph{probes} $1530$, $1524$, $1512$, and $1472 = 1512 - 40$ until $\ALG$ places one on~$H$, see \Cref{fig:easy_phase2} and \Cref{tab:easy_strategy}(b). If the $q$-th probe $J_{20+q}$ ($q \in \{1, 2, 3, 4\}$) is the first to land on~$H$ (Case~$q$), then the machines probed up to Step~$20+q$ reach unreduced loads at least $L = 1973$, $1933$, $1927$, and $1915$ in Cases~1--4, respectively: $H$ reaches $443 + 1530 = 1973$ in Case~1; a light machine reaches $403 + 1530 = 1933$ while $H$ reaches $443 + 1524 = 1967$ in Case~2; two light machines reach $1933$ and $403 + 1524 = 1927$ while $H$ reaches $443 + 1512 = 1955$ in Case~3; and three light machines reach $1933$, $1927$, and $403 + 1512 = 1915$ while $H$ reaches $443 + 1472 = 1915$ in Case~4. The adversary then reduces every probed machine to~$L$ and completes layer~$5$ with $4 - q$ jobs of size $L - 403$ ($1570$, $1530$, or $1524$) on the remaining light machines, bringing all four reduced loads to~$L$. Placing two jobs of layer~$5$ on one machine gives load at least $\min(443 + 1530 + 1570,\, 403 + 2 \times 1570) = 3543 > c_0 \cdot 1985 \approx 3438.13$ in Case~1 ($\OPT \le 1985$, Row~8) and $403 + 1512 + 1472 = 3387 > c_0 \cdot 1945 \approx 3368.85$ in Cases~2--4 ($\OPT \le 1945$, Row~9), so the four jobs of layer~$5$ go to distinct machines.

\input{tikz/fig_simple_endgame}

\paragraph{Phase 3: a linear endgame (Steps 25--29).}
From the common load $l_6 = L$, the adversary finishes as in \Cref{ex:linear}: layer~$6$ consists of three jobs of size~$a$ and one job of size $b > a$, followed by the final job $J_{29} = 2a$, with $(a, b)$ given in \Cref{tab:easy_strategy}(c). Condition~\eqref{eq:linear_layer} for $j = 2, 3$ holds across all four cases since $L + 2a \ge 1915 + 2 \times 4125 = 10165 > c_0 \cdot 5774 \approx 10000.90$; here $\OPT \le 4244 + 1530 = 5774$ in Steps~25--27 comes from pairing the first three layer-$6$ jobs (of size $a \le 4244$) with a layer-$5$ job of size at most~$1530$, two $552$-blocks, and one $552$-block, while the other three layer-$5$ jobs have total size at most $3 \times 1570 = 4710$ on the fourth machine (Row~10). \Cref{tab:easy_strategy}(c) verifies~\eqref{eq:linear_layer} for $j = 4$ (Rows~11, 13, 15, and~17) and~\eqref{eq:linear_final} for $J_{29}$. At Step~29, $\OPT(J_1, \dots, J_{29}) = 2a$: as shown in \Cref{fig:easy_endgame} and in Rows~12, 14, 16, and~18 of \Cref{tab:easy_opt}, $J_{29} = 2a$ runs alone on one machine, two jobs of size~$a$ share a second machine, and the remaining jobs, namely $a$, $b$, the four layer-$5$ jobs, and the three $552$-blocks of \Cref{fig:easy_step20}(c), with $16$ units shifted between two blocks in Case~4, pack onto the other two machines with load at most~$2a$. Reducing $L + b$ to $l_7 = L + a$, the final job $J_{29} = 2a$ forces $\ALG \ge L + 3a \ge c_0 \cdot 2a$ in every case. In the tight branch, Case~$3$, we have $L = 1927$ and $\OPT(J_1, \dots, J_{29}) \le 2a = 8304$, so
\[
  \frac{L + 3a}{2a} \;=\; \frac{1927 + 3 \times 4152}{2 \times 4152} \;=\; \frac{14383}{8304} \;=\; c_0.
\]

\section{Proof of the Optimized Bound (\texorpdfstring{\Cref{thm:opt}}{Theorem 2})}\label{sec:opt}

To prove \Cref{thm:opt} with $c = \tfrac{14350}{8283} \approx 1.73246408$, the adversary again uses $k = 7$ layers of four jobs each ($J_1, \dots, J_{28}$) and a final job $J_{29}$ across three phases, summarized in \Cref{tab:opt_strategy} and verified step by step against explicit offline schedules in \Cref{tab:phase1,tab:phase2,tab:phase3a,tab:phase3b} of \Cref{sec:appendix_tables}. Compared to \Cref{sec:easy}, the strategy makes no load reductions in Phase~1, leaves a single \emph{light} machine below three \emph{heavier} machines in Phase~2 whenever the first probe misses~$H$, and mirrors Phase~2 with \emph{increasing} probes in Phase~3.

\input{tables/tab07_opt_strategy}

\paragraph{Phase 1: a linear prefix without reductions (Steps 1--20).}
The adversary presents the five layers
\[
  (4, 4, 4, 4), \qquad (19, 19, 19, 19), \qquad (59, 59, 59, \mathbf{67}), \qquad (286, 286, 286, 286), \qquad (800, 800, 800, \mathbf{918}).
\]
Because the load $23 + 67 = 90$ of the machine receiving $J_{12}$ is not reduced, $\ALG$ reaches $(90, 82, 82, 82)$ after layer~$2$ and $(376, 368, 368, 368)$ after layer~$3$. Evaluating condition~\eqref{eq:linear_layer} at the minimum machine loads $\underline{l}_0 = 0$, $\underline{l}_1 = 4$, $\underline{l}_2 = 23$, $\underline{l}_3 = 82$, and $\underline{l}_4 = 368$ at the start of the layers, as in \Cref{tab:opt_strategy}(a) and \Cref{tab:phase1} in \Cref{sec:appendix_tables}, forces $\ALG$ to place the four jobs of each layer on distinct machines. In layer~$4$, the fourth job $J_{20} = 918$ must go to the machine that did not receive a job of size~$800$, which we call the \emph{heavy machine}~$H$. Depending on whether $H$ is the machine of $J_{12}$, $\ALG$ ends Phase~1 in one of the two unreduced states
\[
  S_1 = (\mathbf{1294}, 1168, 1168, 1168) \qquad \text{or} \qquad S_2 = (\mathbf{1286}, \mathbf{1176}, 1168, 1168),
\]
where the first entry is the load of~$H$. We refer to the light machines of load~$1168$ as the \emph{$1168$-machines}, and in state~$S_2$ to the machine of load~$1176$ as the \emph{$1176$-machine}. Here $\OPT(J_1, \dots, J_{20}) \le 1204$ (\Cref{tab:phase1}, Step~20). Moreover, the twenty jobs of Phase~1 partition into two blocks of load~$2399$ each, namely $918 + 800 + 2 \times 286 + 59 + 2 \times 19 + 3 \times 4$ and $2 \times 800 + 2 \times 286 + 67 + 2 \times 59 + 2 \times 19 + 4$, which we use to bound $\OPT$ in Steps~25--27 below.

\paragraph{Phase 2: a branching layer with state-dependent probes (Steps 21--24).}
The adversary presents decreasing probes as long as $\ALG$ places them on light machines, and completes layer~$5$ once a probe lands on~$H$, see \Cref{tab:opt_strategy}(b). The first probe is $J_{21} = 4620$.
\begin{itemize}
  \item If $\ALG$ places $J_{21}$ on~$H$ (Case~1), three jobs of size~$4730$ complete the layer on the three light machines, and the adversary reduces all four loads to the common load $L := 5898$.
  \item Otherwise, the subsequent probes depend on where $J_{21}$ was placed: they are $J_{22} = 4578$ and $J_{23} = 4535$ if $J_{21}$ went to a $1168$-machine, and $J_{22} = 4586$ and $J_{23} = 4546$ if $J_{21}$ went to the $1176$-machine of~$S_2$. If $\ALG$ places $J_{22}$ on~$H$, two jobs of size~$4666$ (Case~2a) or~$4674$ (Case~2b) complete the layer; if $\ALG$ places $J_{23}$ on~$H$, a single job of size~$4620$ (Case~3a) or~$4628$ (Case~3b) completes the layer.
  \item If all three probes go to light machines (Cases~4a--4d), then $J_{24}$ must go to~$H$, with size $4452$ in Case~4a (from~$S_1$), $4460$ in Case~4b ($J_{23}$ on the $1176$-machine), and $4468$ in Cases~4c and~4d ($J_{22}$ or $J_{21}$ on the $1176$-machine).
\end{itemize}
Placing two jobs of layer~$5$ on one machine is an immediate loss: during Steps~21--23, $\OPT \le 4730$ and any such machine reaches load at least $1168 + 4578 + 4535 = 10281 > c \cdot 4730 \approx 8194.56$; at Step~24, its load reaches at least $1168 + 2 \times 4730 = 10628 > c \cdot 5902 \approx 10225.00$ in Case~1, and at least $1168 + 4535 + 4452 = 10155 > c \cdot 5840 \approx 10117.59$ in Cases~2a--4d (\Cref{tab:phase2} in \Cref{sec:appendix_tables}). Hence the four jobs of layer~$5$ occupy distinct machines. Whereas \Cref{sec:easy} chooses completion jobs equal to the last light-machine probe to equalize all four reduced loads to~$L$, here the completion jobs ($4666$, $4674$, $4620$, $4628$) are chosen slightly larger than the preceding probes so that every unprobed light machine is lifted to~$h$, leaving only the machine of the last non-$H$ probe at the lighter reduced load $l = h - g$. Specifically, the unreduced loads after Step~24 are $(5872, 5834, 5834, 5788)$ or $(5864, 5842, 5834, 5788)$ in Case~2a; $(5872, 5842, 5842, 5796)$ in Case~2b; $(5829, 5788, 5788, 5746)$, $(5821, 5796, 5788, 5746)$, or $(5821, 5788, 5788, 5754)$ in Case~3a; $(5832, 5796, 5796, 5754)$ in Case~3b; and $(5788, 5746, 5746, 5703)$, $(5788, 5746, 5746, 5711)$, $(5788, 5754, 5754, 5703)$, and $(5796, 5754, 5754, 5714)$ in Cases~4a--4d (\Cref{tab:phase2}). In each case, the adversary reduces the machine of the last non-$H$ probe to~$l$ and the other three machines to~$h$, reaching the reduced state $(h, h, h, l)$ in \Cref{tab:opt_strategy}(b) where the \emph{light} machine sits $g := h - l \in [35, 51]$ units below the three \emph{heavier} machines.

\paragraph{Phase 3: a linear endgame or a mirrored branching layer (Steps 25--29).}
In Case~1, the adversary finishes with a linear endgame as in \Cref{sec:easy}: three jobs of size $a = 12652$, one job of size $b = 14348$, and the final job $J_{29}$ of size $25304 = 2a$. Condition~\eqref{eq:linear_layer} holds since $L + 2a = 31202 > c \cdot 17272 \approx 29923.12$ and $L + a + b = 32898 > c \cdot 18982 \approx 32885.63$, and condition~\eqref{eq:linear_final} holds since $L + 3a = 43854 > c \cdot 25305 \approx 43840.00$ (\Cref{tab:phase2}).

In Cases~2a--4d, the light machine starts $g = h - l$ units below the three heavier machines. Mirroring Phase~2, the adversary presents increasing probes $s_1 < s_2 < s_3 = s_1 + g$ until $\ALG$ places one on the light machine, see \Cref{tab:opt_strategy}(c):
\begin{itemize}[nosep]
  \item[(i)] If $\ALG$ places $J_{25} = s_1$ on the light machine, two jobs of size $s_1 - g$ follow on two heavier machines, bringing three machines to load at least $E = l + s_1 = h + (s_1 - g)$.
  \item[(ii)] If $s_1$ goes to a heavier machine and $J_{26} = s_2$ to the light machine, one job of size $s_2 - g$ brings a second heavier machine to load at least $E = l + s_2 = h + (s_2 - g)$, while the machine of $s_1$ has load at least $h + s_1 = l + s_3 > E$.
  \item[(iii)] If $s_1$ and $s_2$ both go to heavier machines, the adversary presents $J_{27} = s_3 = s_1 + g$. Either $\ALG$ places $s_3$ on the light machine (giving it load $l + s_3 = h + s_1 = E$) and the layer-completion job $J_{28} = b > s_3$ goes to the third heavier machine (load $h + b > h + s_3 > E$), or $\ALG$ places $s_3$ on the third heavier machine (load $h + s_3 > E$) and $J_{28} = b$ goes to the light machine (load $l + b > l + s_3 = E$).
\end{itemize}
In each subcase, the fourth job $J_{28} = b > s_3$ completes layer~$6$ on the remaining machine (raising its load above $l + s_3 \ge E$), followed by the final job $J_{29} = f$. Across Cases~2a--4d, every machine has load at least $l \ge 5703$ after Phase~2, and every job presented in Steps~25--27 has size at least $s_1 - g \ge 12321 - 43 = 12278$, attained in Case~4a(i), and at most $s_3 \le 12575$, attained in Case~2b(iii). Placing two jobs of layer~$6$ on one machine is therefore an immediate loss during Steps~26--27: any such machine reaches load at least $5703 + 2 \times 12278 = 30259 > c \cdot 17161 \approx 29730.80$. Here $\OPT \le 12575 + 4586 = 17161$ comes from pairing $J_{25}$ and $J_{26}$ with the two Phase-1 blocks of load~$2399$, pairing $J_{27}$ (if present) with the smallest layer-$5$ job of size at most~$4586$, and placing the remaining three layer-$5$ jobs of total size at most~$13968$ on the fourth machine. At Step~28, placing $J_{28} = b$ on a machine that already holds a job of layer~$6$ raises its load to at least $E + b > c \cdot \OPT(J_1, \dots, J_{28})$ across all $24$ subcases of Cases~2a--4d, as verified for every subcase in \Cref{tab:phase3a,tab:phase3b} of \Cref{sec:appendix_tables}; the tightest subcase is Case~4a(ii), where $E + b = 18048 + 13968 = 32016 > c \cdot 18480 \approx 32015.94$. Thus the four jobs of layer~$6$ go to distinct machines, raising every machine to load at least~$E$ before Step~29. Wherever $\ALG$ places $J_{29} = f$, its makespan reaches at least $E + f \ge c \cdot \OPT(J_1, \dots, J_{29})$, see \Cref{tab:opt_strategy}(c) and \Cref{tab:phase3a,tab:phase3b}. In the tight branch, Case~3a(iii), we have $E = 5788 + 12413 = 18201$ and $\OPT(J_1, \dots, J_{29}) \le f = 24849$, so
\[
  \frac{E + f}{f} \;=\; \frac{43050}{24849} \;=\; \frac{14350}{8283} \;=\; c.
\]

\section{Conclusion}\label{sec:conclusion}

Closing the remaining gap between $\tfrac{14350}{8283} \approx 1.73246$ and $\tfrac{26}{15} \approx 1.73333$ on $m = 4$ machines will require deeper branching trees on the lower bound side, refined online algorithms, or both; in particular, resolving whether $\sqrt{3}$ is optimal in the incremental model (\Cref{sec:linear}) would separate branching from linear adversaries on four machines. More broadly, branching layers with non-monotone probes can be combined with multi-machine linear prefixes and may yield improved lower bounds for $m \ge 5$ and for large~$m$, where the best known lower bound of $1.88$~\cite{Rudin2001} is still proved by a linear adversary.

\bibliographystyle{plainnat}
\bibliography{references}

\clearpage
\appendix
\crefalias{section}{appendix}
\section{Step-by-Step Verification Tables for \texorpdfstring{\Cref{thm:opt}}{Theorem 2}}\label{sec:appendix_tables}

\Cref{tab:phase1,tab:phase2,tab:phase3a,tab:phase3b} provide the complete step-by-step verification of the adversary strategy for \Cref{thm:opt} ($c = \tfrac{14350}{8283} \approx 1.73246408$). \Cref{tab:phase1} covers Phase~1 (Steps~1--20), \Cref{tab:phase2} covers Phase~2 (Steps~21--24) together with the linear endgame of Case~1 (Steps~25--29), and \Cref{tab:phase3a,tab:phase3b} cover Phase~3 (Steps~25--29) for Cases~2a--3b and Cases~4a--4d across Subcases~(i)--(iii).
In Phases~1 and~2 (Steps~1--24 of \Cref{tab:phase1,tab:phase2}), the tables track the exact unreduced machine loads of $\ALG$, recording at Step~24 the resulting unreduced load tuples before they are reduced to $L = 5898$ in Case~1 or to $(h, h, h, l)$ in Cases~2a--4d, as in \Cref{tab:opt_strategy}(b). In Phase~3 (Steps~25--29 of \Cref{tab:phase2,tab:phase3a,tab:phase3b}), the analysis starts from these reduced Phase-2 loads $L = 5898$ and $(h, h, h, l)$, and at Steps~28--29 the column $\ALG \ge$ records $E + b$ and $E + f$ from the reduced load~$E$ in \Cref{tab:opt_strategy}(c).
In each row of Steps~1--28, the column \emph{Valid} (or \emph{Valid / Transition}) lists the machine loads from the start of the current layer on which $\ALG$ can place $J_t$ without losing immediately, together with the resulting branch transition. Placing $J_t$ on any machine of larger current load (in particular, on a machine that already received a job of the current layer, or on any machine at Step~29) raises the makespan of $\ALG$ to at least the value in the $\ALG \ge$ column. Finally, the last column exhibits an explicit four-machine schedule ($\mathbf{load}:\text{jobs}$) of all jobs presented up to that step whose maximum load witnesses the stated upper bound on $\OPT$ and confirms that $\ALG \ge c \cdot \OPT$.

\input{tables/tab02_opt_phase1}

\input{tables/tab03_opt_phase2}

\input{tables/tab04_opt_phase3a}

\input{tables/tab05_opt_phase3b}

\end{document}

%% file: tikz/fig_linear_adversary.tex
\begin{figure}[!tb]
\centering
\begin{tikzpicture}[
  x=1cm, y=1cm,
  job/.style={draw=black, semithick},
  cut/.style={draw=black!70, densely dashed, semithick, fill=jobPaleBlue},
  ghost/.style={draw=red!75!black, densely dashed, thick, fill=red!8},
  boundline/.style={dashed, line width=1.2pt, black},
  lbl/.style={font=\scriptsize, inner sep=0.5pt},
  wlbl/.style={lbl, text=white},
  slbl/.style={font=\fontsize{5.0}{5.5}\selectfont, inner sep=0pt},
  panelcap/.style={font=\footnotesize, anchor=north, align=center},
  brc/.style={decorate, decoration={brace, amplitude=3pt}, semithick},
]
  \begin{scope}
    \def\colw{0.9}   
    \def\hLoad{1.35} 
    \def\hJob{0.95}  
    \pgfmathsetmacro{\hTop}{\hLoad+\hJob}  
    \foreach \mi in {0,...,3} {
      \filldraw[job, fill=jobLightGray] ({\mi*\colw},0) rectangle ({(\mi+1)*\colw},\hLoad);
      \node[lbl] at ({(\mi+0.5)*\colw},{0.5*\hLoad}) {$l_i$};
    }
    \foreach \mi/\extra in {1/0.24, 2/0.48, 3/0.72} {
      \draw[cut] ({\mi*\colw},\hTop) rectangle ({(\mi+1)*\colw},{\hTop+\extra});
    }
    \foreach \mi/\jj in {0/1, 1/2, 2/3, 3/4} {
      \filldraw[job, fill=jobMedBlue] ({\mi*\colw},\hLoad) rectangle ({(\mi+1)*\colw},\hTop);
      \node[wlbl] at ({(\mi+0.5)*\colw},{\hLoad+0.5*\hJob}) {$J_{4i+\jj}$};
    }
    \filldraw[ghost] (0,\hTop) rectangle (\colw,{\hTop+1.15});
    \node[lbl, text=red!75!black] at ({0.5*\colw},{\hTop+0.575}) {$J_{4i+j}$};
    \draw[brc, red!75!black] (-0.12,0) -- (-0.12,{\hTop+1.15});
    \node[lbl, text=red!75!black, anchor=east] at (-0.27,{0.5*(\hTop+1.15)}) {$\ge c\cdot\OPT$};
    \draw[boundline] (-0.08,\hTop) -- ({4*\colw+0.08},\hTop);
    \node[lbl, anchor=west] at ({4*\colw+0.12},\hTop) {$l_{i+1} = l_i + p$};
    \node[lbl, align=center, text=black!85] (cutlbl) at ({2.15*\colw},{\hTop+1.44}) {cut off by\\the adversary};
    \draw[-{Stealth[length=1.5mm]}, black!80, semithick] ([xshift=-6pt]cutlbl.south) -- ({1.5*\colw},{\hTop+0.18});
    \draw[-{Stealth[length=1.5mm]}, black!80, semithick] (cutlbl.south) -- ({2.5*\colw},{\hTop+0.40});
    \draw[-{Stealth[length=1.5mm]}, black!80, semithick] ([xshift=6pt]cutlbl.south) -- ({3.5*\colw},{\hTop+0.62});
    \drawMachineTray{0}{0}{\colw}
    \node[panelcap] at ({2*\colw},-0.22) {(a) Layer $i$};
  \end{scope}

  \begin{scope}[shift={(6.15,0)}]
    \def\colw{0.72}
    \def\hone{0.19}                             
    \def\hx{0.4867}                             
    \pgfmathsetmacro{\ha}{2*\hx+3*\hone}        
    \pgfmathsetmacro{\hb}{\ha+2*\hone}          
    \pgfmathsetmacro{\hL}{\hone+\hx}            
    \pgfmathsetmacro{\hla}{\hL+\ha}             
    \foreach \mi in {0,...,3} {
      \filldraw[job, fill=jobLightGray] ({\mi*\colw},0) rectangle ({(\mi+1)*\colw},\hone);
      \node[slbl] at ({(\mi+0.5)*\colw},{0.5*\hone}) {$1$};
      \filldraw[job, fill=jobPaleBlue] ({\mi*\colw},\hone) rectangle ({(\mi+1)*\colw},\hL);
      \node[lbl] at ({(\mi+0.5)*\colw},{\hone+0.5*\hx}) {$x$};
    }
    \foreach \mi in {0, 1, 2} {
      \filldraw[job, fill=jobMedBlue] ({\mi*\colw},\hL) rectangle ({(\mi+1)*\colw},\hla);
      \node[wlbl] at ({(\mi+0.5)*\colw},{\hL+0.5*\ha}) {$a$};
    }
    \filldraw[job, fill=jobLightBlue] ({3*\colw},\hL) rectangle ({4*\colw},\hla);
    \node[lbl] at ({3.5*\colw},{\hL+0.5*\ha}) {$a$};
    \draw[cut, fill=jobLightBlue!35] ({3*\colw},\hla) rectangle ({4*\colw},{\hla+2*\hone});
    \node[lbl, text=black!85] at ({3.5*\colw},{\hla+\hone}) {$b - a$};
    \draw[boundline] (-0.08,\hla) -- ({4*\colw+0.08},\hla);
    \node[lbl, anchor=west] at ({4*\colw+0.12},\hla) {$l_3 = L + a$};
    \filldraw[job, fill=jobOrange] (0,\hla) rectangle (\colw,{\hla+2*\ha});
    \node[wlbl] at ({0.5*\colw},{\hla+\ha}) {$2a$};
    \node[lbl, anchor=west] at ({\colw+0.12},{\hla+\ha}) {final job};
    \draw[boundline] (-0.08,{\hla+2*\ha}) -- ({4*\colw+0.08},{\hla+2*\ha});
    \node[lbl, anchor=west] at ({4*\colw+0.12},{\hla+2*\ha}) {$c \cdot \OPT$};
    \draw[brc] ({4*\colw+0.1},\hL) -- ({4*\colw+0.1},0);
    \node[lbl, anchor=west] at ({4*\colw+0.25},{0.5*\hL}) {$L = 1 + x$};
    \drawMachineTray{0}{0}{\colw}
    \node[panelcap] at ({2*\colw},-0.22) {(b) $\ALG = L + 3a$};
  \end{scope}

  \begin{scope}[shift={(11.0,0)}]
    \def\colw{0.72}
    \def\hone{0.19}
    \def\hx{0.4867}
    \pgfmathsetmacro{\ha}{2*\hx+3*\hone}
    \pgfmathsetmacro{\hb}{\ha+2*\hone}
    \filldraw[job, fill=jobOrange] (0,0) rectangle (\colw,{2*\ha});
    \node[wlbl] at ({0.5*\colw},\ha) {$2a$};
    \foreach \row in {0,1} {
      \filldraw[job, fill=jobMedBlue] (\colw,{\row*\ha}) rectangle ({2*\colw},{(\row+1)*\ha});
      \node[wlbl] at ({1.5*\colw},{(\row+0.5)*\ha}) {$a$};
    }
    \foreach \row in {0,1,2} {
      \filldraw[job, fill=jobLightGray] ({2*\colw},{\row*\hone}) rectangle ({3*\colw},{(\row+1)*\hone});
      \node[slbl] at ({2.5*\colw},{(\row+0.5)*\hone}) {$1$};
    }
    \foreach \row in {0,1} {
      \filldraw[job, fill=jobPaleBlue] ({2*\colw},{3*\hone+\row*\hx}) rectangle ({3*\colw},{3*\hone+(\row+1)*\hx});
      \node[lbl] at ({2.5*\colw},{3*\hone+(\row+0.5)*\hx}) {$x$};
    }
    \filldraw[job, fill=jobMedBlue] ({2*\colw},{3*\hone+2*\hx}) rectangle ({3*\colw},{2*\ha});
    \node[wlbl] at ({2.5*\colw},{3*\hone+2*\hx+0.5*\ha}) {$a$};
    \filldraw[job, fill=jobLightGray] ({3*\colw},0) rectangle ({4*\colw},\hone);
    \node[slbl] at ({3.5*\colw},{0.5*\hone}) {$1$};
    \foreach \row in {0,1} {
      \filldraw[job, fill=jobPaleBlue] ({3*\colw},{\hone+\row*\hx}) rectangle ({4*\colw},{\hone+(\row+1)*\hx});
      \node[lbl] at ({3.5*\colw},{\hone+(\row+0.5)*\hx}) {$x$};
    }
    \filldraw[job, fill=jobLightBlue] ({3*\colw},{\hone+2*\hx}) rectangle ({4*\colw},{2*\ha});
    \node[lbl] at ({3.5*\colw},{\hone+2*\hx+0.5*\hb}) {$b$};
    \draw[boundline] (-0.08,{2*\ha}) -- ({4*\colw+0.08},{2*\ha});
    \node[lbl, anchor=west] at ({4*\colw+0.12},{2*\ha}) {$\OPT$};
    \drawMachineTray{0}{0}{\colw}
    \node[panelcap] at ({2*\colw},-0.22) {(c) $\OPT = 2a$};
  \end{scope}
\end{tikzpicture}
\caption{A linear adversary. (a)~Layer~$i$ (schematic): all machines start with load~$l_i$. If $\ALG$ placed a second job $J_{4i+j}$ of the layer on a machine (red, dashed), this machine would reach load $l_i + p + p_{4i+j} \ge c \cdot \OPT$ by~\eqref{eq:linear_layer}. Hence, the four jobs of the layer go to distinct machines, and the adversary cuts off the parts of $J_{4i+2}, J_{4i+3}, J_{4i+4}$ above $l_i + p$ (light blue, dashed), leaving the retained part of height $p = p_{4i+1}$ (dark blue) on every machine so that each ends the layer with load $l_{i+1} = l_i + p$. (b)~The instance of \Cref{ex:linear} (drawn to scale): jobs of size~$1$ (gray) and~$x$ lift all loads to $L = 1 + x$; layer~$2$ consists of three jobs of size~$a$ and one larger job $J_{12}$ of size $b = a + 2$, whose excess $b - a = 2$ above $l_3 = L + a$ is cut off. The final job of size~$2a$ forces $\ALG \ge L + 3a = c \cdot 2a$. (c)~An optimal offline schedule of the same $13$ jobs. Nothing is cut off for $\OPT$: load reductions apply only to the loads of $\ALG$, so $\OPT$ schedules every job at its full size (here $J_{12}$ with size~$b$). The final job runs alone, two jobs of size~$a$ share a machine, and the remaining jobs fill the other two machines, so $\OPT = 2a$.}
\label{fig:linear_adversary}
\end{figure}
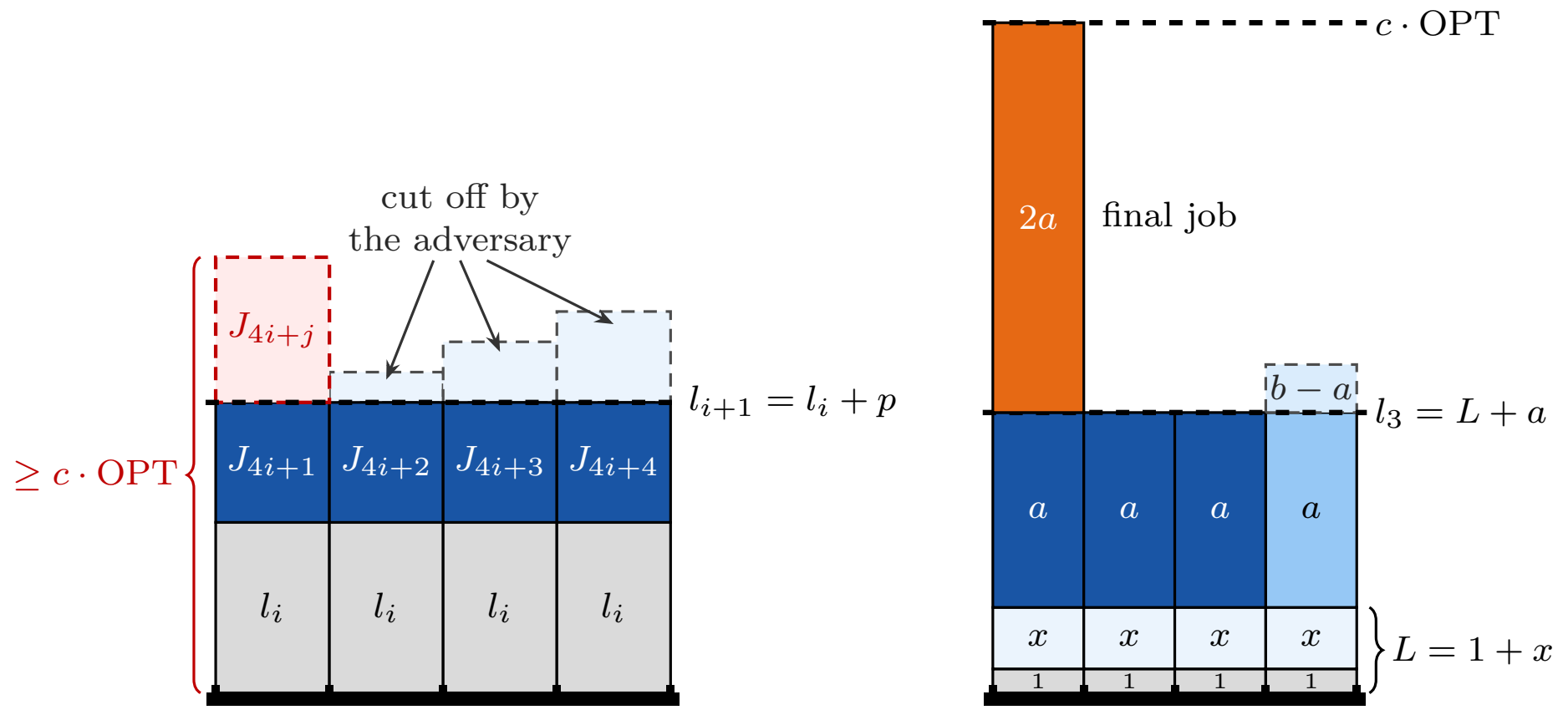

%% file: tables/tab06_easy_strategy.tex
\begin{table}[!tb]
\centering
\footnotesize
\setlength{\fboxsep}{1.1pt}
\renewcommand{\arraystretch}{1.12}
\setlength{\tabcolsep}{2.5pt}
\def\rowhd#1{\begin{tabular}[t]{@{}r@{}}#1\\(Row)\end{tabular}}
\caption{The strategy of the adversary for \Cref{thm:easy}, where $c_0 = \frac{14383}{8304}$ and $\OPT_t := \OPT(J_1, \dots, J_t)$. In each condition, the three columns give an upper bound on~$\OPT_t$ together with the row of \Cref{tab:easy_opt} that witnesses it, $c_0$ times this bound, and the makespan of $\ALG$ if it places two jobs of the same layer on one machine or, for condition~\eqref{eq:linear_final}, after the final job. (a)~Phase~1: the enlarged last jobs of layers~$2$ and~$4$ (bold) need a separate condition for $j = 4$. (b)~Phase~2: Case~$q$ occurs if $J_{20+q}$ (orange) is the first probe that $\ALG$ places on~$H$. The earlier probes (light blue) went to light machines, listed in the order in which they receive their layer-$5$ job, and the remaining jobs complete the layer; for Cases~2--4, $3387 = 403 + 1512 + 1472$ is a common bound. (c)~Phase~3: the linear endgame of \Cref{ex:linear} from the common load~$L$. Condition~\eqref{eq:linear_layer} forces the jobs $a, a, a, b$ of layer~$6$ onto distinct machines, after which the final job $2a$ raises the makespan to $L + 3a$; the bold entries (Case~3) form the tight inequality, which determines~$c_0$.}
\label{tab:easy_strategy}

\begin{tabular*}{\linewidth}{@{\extracolsep{\fill}}c r rrrr rrr rrr@{}}
\parthead{12}{\textbf{(a)}\enspace Phase~1: linear prefix, layers $i = 0, \dots, 4$ (Steps 1--20)}
\toprule
 & & \multicolumn{4}{c}{Jobs of layer $i$} & \multicolumn{3}{c}{Condition~\eqref{eq:linear_layer}, $j = 2, 3$} & \multicolumn{3}{c}{Condition~\eqref{eq:linear_layer}, $j = 4$} \\
\cmidrule(lr){3-6}\cmidrule(lr){7-9}\cmidrule(l){10-12}
$i$ & $l_i$ & $p_{4i+1}$ & $p_{4i+2}$ & $p_{4i+3}$ & $p_{4i+4}$
    & \rowhd{$\OPT_{4i+3} \le$} & $c_0 \cdot \OPT_{4i+3}$ & $l_i + 2p$
    & \rowhd{$\OPT_{4i+4} \le$} & $c_0 \cdot \OPT_{4i+4}$ & $l_i + p + p_{4i+4}$ \\
\midrule
0 &   0 &   2 &   2 &   2 &            2 &   2\,(1) &   3.46 &   4 &   2\,(1) &   3.46 &   4 \\
1 &   2 &   6 &   6 &   6 &            6 &   8\,(2) &  13.86 &  14 &   8\,(2) &  13.86 &  14 \\
2 &   8 &  20 &  20 &  20 & \textbf{24}  &  24\,(3) &  41.57 &  48 &  30\,(4) &  51.96 &  52 \\
3 &  28 &  99 &  99 &  99 &           99 & 129\,(5) & 223.44 & 226 & 129\,(5) & 223.44 & 226 \\
4 & 127 & 276 & 276 & 276 & \textbf{316} & 392\,(6) & 678.97 & 679 & 415\,(7) & 718.80 & 719 \\
\bottomrule
\end{tabular*}

\vspace{5pt}
\begin{tabular*}{\linewidth}{@{\extracolsep{\fill}}c cccc rrr@{\hspace{10pt}}rrrr r@{}}
\parthead{13}{\textbf{(b)}\enspace Phase~2: branching layer~$5$ (Steps 21--24), starting from the loads $(443, 403, 403, 403)$}
\toprule
 & \multicolumn{4}{c}{Jobs of layer $5$} & \multicolumn{3}{c}{Two jobs on one machine} & \multicolumn{4}{c}{Unreduced loads after Step~24} & Reduced \\
\cmidrule(lr){2-5}\cmidrule(lr){6-8}\cmidrule(lr){9-12}\cmidrule(l){13-13}
Case & $p_{21}$ & $p_{22}$ & $p_{23}$ & $p_{24}$
     & \rowhd{$\OPT_{24} \le$} & $c_0 \cdot \OPT_{24}$ & load $\ge$
     & $H$ & light~$1$ & light~$2$ & light~$3$ & $L$ \\
\midrule
1 & \onH{1530}   & \compl{1570} & \compl{1570} & \compl{1570} & 1985\,(8) & 3438.13 & 3543 & 1973 & 1973 & 1973 & 1973 & 1973 \\
2 & \probe{1530} & \onH{1524}   & \compl{1530} & \compl{1530} & 1945\,(9) & 3368.85 & 3387 & 1967 & 1933 & 1933 & 1933 & 1933 \\
3 & \probe{1530} & \probe{1524} & \onH{1512}   & \compl{1524} & 1945\,(9) & 3368.85 & 3387 & 1955 & 1933 & 1927 & 1927 & 1927 \\
4 & \probe{1530} & \probe{1524} & \probe{1512} & \onH{1472}   & 1945\,(9) & 3368.85 & 3387 & 1915 & 1933 & 1927 & 1915 & 1915 \\
\bottomrule
\end{tabular*}

\vspace{5pt}
\begin{tabular*}{\linewidth}{@{\extracolsep{\fill}}c r@{\hspace{7pt}}r@{\hspace{7pt}}r@{\hspace{9pt}}rrr@{\hspace{9pt}}rrr@{\hspace{9pt}}rrr@{}}
\parthead{13}{\textbf{(c)}\enspace Phase~3: linear endgame with jobs of sizes $a, a, a, b, 2a$ (Steps 25--29), starting from the common load~$L$}
\toprule
 & & & & \multicolumn{3}{c}{Condition~\eqref{eq:linear_layer}, $j = 2, 3$} & \multicolumn{3}{c}{Condition~\eqref{eq:linear_layer}, $j = 4$} & \multicolumn{3}{c}{Condition~\eqref{eq:linear_final}} \\
\cmidrule(lr){5-7}\cmidrule(lr){8-10}\cmidrule(l){11-13}
Case & $L$ & $a$ & $b$
     & \rowhd{$\OPT_{27} \le$} & $c_0 \cdot \OPT_{27}$ & $L + 2a$
     & \rowhd{$\OPT_{28} \le$} & $c_0 \cdot \OPT_{28}$ & $L + a + b$
     & \rowhd{$\OPT_{29} \le$} & $c_0 \cdot \OPT_{29}$ & $L + 3a$ \\
\midrule
1 & 1973 & 4244 & 4836 & 5774\,(10) & 10000.90 & 10461 & 6366\,(11) & 11026.27 & 11053 & 8488\,(12) & 14701.70 & 14705 \\
2 & 1933 & 4164 & 4722 & 5774\,(10) & 10000.90 & 10261 & 6246\,(13) & 10818.43 & 10819 & 8328\,(14) & 14424.57 & 14425 \\
3 & 1927 & 4152 & 4710 & 5774\,(10) & 10000.90 & 10231 & 6228\,(15) & 10787.25 & 10789 & 8304\,(16) & \textbf{14383.00} & \textbf{14383} \\
4 & 1915 & 4125 & 4680 & 5774\,(10) & 10000.90 & 10165 & 6189\,(17) & 10719.70 & 10720 & 8250\,(18) & 14289.47 & 14290 \\
\bottomrule
\end{tabular*}
\end{table}

%% file: tikz/fig_simple_step20.tex
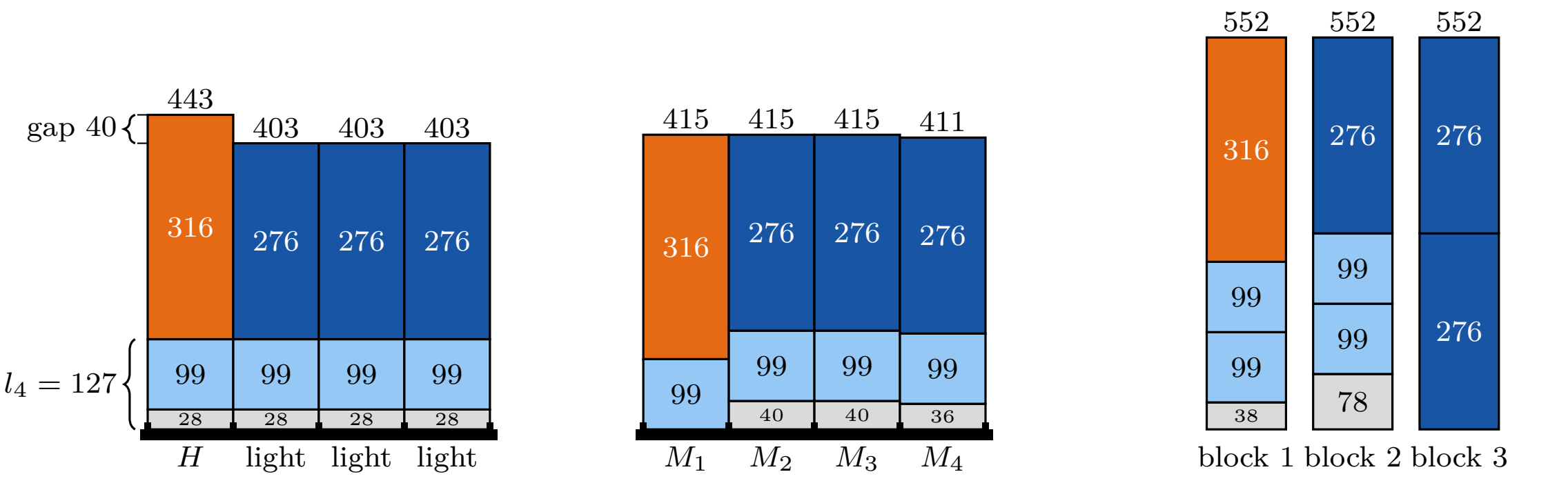
\begin{figure}[!tb]
\centering
\begin{tikzpicture}[
  x=1cm, y=1cm,
  job/.style={draw=black, semithick},
  lbl/.style={font=\scriptsize, inner sep=0.5pt},
  wlbl/.style={lbl, text=white},
  slbl/.style={font=\fontsize{5.0}{5.5}\selectfont, inner sep=0pt},
  toplbl/.style={font=\scriptsize, anchor=south, inner sep=1.5pt},
  panelcap/.style={font=\footnotesize, anchor=north, align=center},
  brc/.style={decorate, decoration={brace, amplitude=3pt}, semithick},
]
  \def\colw{0.82}
  \def\JBox#1#2#3#4#5#6{%
    \filldraw[job, fill=#4] ({#1*\colw},#2) rectangle ({(#1+1)*\colw},#3);
    \node[#5] at ({(#1+0.5)*\colw},{0.5*(#2+#3)}) {$#6$};
  }
  \begin{scope}
    \begin{scope}[y=0.0068cm]
      \foreach \mi in {0,...,3} {
        \JBox{\mi}{0}{28}{jobLightGray}{slbl}{28}
        \JBox{\mi}{28}{127}{jobLightBlue}{lbl}{99}
      }
      \JBox{0}{127}{443}{jobOrange}{wlbl}{316}
      \node[toplbl] at ({0.5*\colw},443) {$443$};
      \foreach \mi in {1,2,3} {
        \JBox{\mi}{127}{403}{jobMedBlue}{wlbl}{276}
        \node[toplbl] at ({(\mi+0.5)*\colw},403) {$403$};
      }
      \draw[densely dashed, thin] (-0.10,403) -- (0,403);
      \draw[densely dashed, thin] (-0.10,443) -- (0,443);
      \draw[brc] (-0.12,0) -- (-0.12,127);
      \node[lbl, anchor=east] at (-0.27,63.5) {$l_4 = 127$};
      \draw[brc] (-0.12,403) -- (-0.12,443);
      \node[lbl, anchor=east] at (-0.27,423) {gap $40$};
    \end{scope}
    \drawMachineTray{0}{0}{\colw}
    \node[lbl, anchor=north] at ({0.5*\colw},-0.14) {$H$};
    \foreach \mi in {1,2,3} {\node[lbl, anchor=north] at ({(\mi+0.5)*\colw},-0.14) {light};}
    \node[panelcap] at ({2*\colw},-0.46) {(a) $\ALG$ after Step~20};
  \end{scope}

  \begin{scope}[shift={(4.75,0)}]
    \begin{scope}[y=0.0068cm]
      \JBox{0}{0}{99}{jobLightBlue}{lbl}{99}
      \JBox{0}{99}{415}{jobOrange}{wlbl}{316}
      \foreach \mi in {1, 2} {
        \JBox{\mi}{0}{40}{jobLightGray}{slbl}{40}
        \JBox{\mi}{40}{139}{jobLightBlue}{lbl}{99}
        \JBox{\mi}{139}{415}{jobMedBlue}{wlbl}{276}
      }
      \JBox{3}{0}{36}{jobLightGray}{slbl}{36}
      \JBox{3}{36}{135}{jobLightBlue}{lbl}{99}
      \JBox{3}{135}{411}{jobMedBlue}{wlbl}{276}
      \foreach \mi in {0, 1, 2} {
        \node[toplbl] at ({(\mi+0.5)*\colw},415) {$415$};
      }
      \node[toplbl] at ({3.5*\colw},411) {$411$};
    \end{scope}
    \drawMachineTray{0}{0}{\colw}
    \foreach \mi/\mnum in {0/1, 1/2, 2/3, 3/4} {\node[lbl, anchor=north] at ({(\mi+0.5)*\colw},-0.14) {$M_{\mnum}$};}
    \node[panelcap] at ({2*\colw},-0.46) {(b) Offline schedule, $\OPT \le 415$};
  \end{scope}

  \begin{scope}[shift={(10.15,0)}]
    \def\bw{0.76}
    \def\bstep{1.02}
    \def\BBox#1#2#3#4#5#6{%
      \filldraw[job, fill=#4] ({#1*\bstep},#2) rectangle ({#1*\bstep+\bw},#3);
      \node[#5] at ({#1*\bstep+0.5*\bw},{0.5*(#2+#3)}) {$#6$};
    }
    \begin{scope}[y=0.0068cm]
      \BBox{0}{0}{38}{jobLightGray}{slbl}{38}
      \BBox{0}{38}{137}{jobLightBlue}{lbl}{99}
      \BBox{0}{137}{236}{jobLightBlue}{lbl}{99}
      \BBox{0}{236}{552}{jobOrange}{wlbl}{316}
      \BBox{1}{0}{78}{jobLightGray}{lbl}{78}
      \BBox{1}{78}{177}{jobLightBlue}{lbl}{99}
      \BBox{1}{177}{276}{jobLightBlue}{lbl}{99}
      \BBox{1}{276}{552}{jobMedBlue}{wlbl}{276}
      \BBox{2}{0}{276}{jobMedBlue}{wlbl}{276}
      \BBox{2}{276}{552}{jobMedBlue}{wlbl}{276}
      \foreach \mi in {0,1,2} {\node[toplbl] at ({\mi*\bstep+0.5*\bw},552) {$552$};}
    \end{scope}
    \foreach \mi/\bi in {0/1, 1/2, 2/3} {\node[lbl, anchor=north] at ({\mi*\bstep+0.5*\bw},-0.14) {block~$\bi$};}
    \node[panelcap] at ({\bstep+0.5*\bw},-0.46) {(c) Three blocks of size $552$};
  \end{scope}
\end{tikzpicture}
\caption{The state after Phase~1 (Step~20). (a)~$\ALG$: every machine received one job from each of layers $0$--$3$, giving the common load $l_4 = 127$; in layer~$4$, the three light machines received $276$ and the heavy machine~$H$ received $J_{20} = 316$. (b)~The four-machine offline schedule of loads $(415, 415, 415, 411)$ from Row~7 of \Cref{tab:easy_opt}, to which Phase~2 adds one layer-$5$ job per machine. (c)~The same twenty jobs partitioned into three blocks of load~$552$, which Phase~3 uses in \Cref{fig:easy_endgame}. Each gray box aggregates the jobs from layers $0$--$2$ on that machine or block.}
\label{fig:easy_step20}
\end{figure}

%% file: tables/tab01_easy_opt.tex
\begin{table}[!tb]
\centering
\scriptsize
\setlength{\tabcolsep}{2.5pt}
\renewcommand{\arraystretch}{1.02}
\caption{Complete verification of all 18 offline upper bounds ($\OPT$) for \Cref{thm:easy} ($c_0 = \frac{14383}{8304} > \sqrt{3}$). Bracketed numbers~$[B]$ denote composite prefix blocks rather than single jobs: $[30]$ and $[28]$ in Row~5 are the Row~4 bundles; $[415]$ and $[411]$ in Rows~8--9 are the Row~7 bundles; and $[552]$ in Rows~10--18 are the three Phase-$1$ blocks of \Cref{fig:easy_step20}(c). Rows~10--14 and~16 use the three $[552]$ blocks unsplit; Row~15 moves a single~$6$ out of the first block, leaving $[546] = [552 - 6]$; Row~17 moves $20$ from the first block and $2 \times 6 + 2 = 14$ from the second onto the first machine, leaving $[532] = [552 - 20]$ and $[538] = [552 - 14]$; and Row~18 moves $2 \times 6 + 2 \times 2 = 16$ between the first two blocks, giving $[568] = [552 + 16]$ and $[536] = [552 - 16]$.}
\label{tab:easy_opt}
\begin{tabular*}{\linewidth}{@{\extracolsep{\fill}}r l l p{0.642\linewidth}@{}}
\toprule
\textbf{Row} & \textbf{Scope} & \textbf{$\OPT$ Bound} & \textbf{Offline 4-Machine Schedule, $\mathbf{load}:\text{jobs and }\text{[prefix blocks]}$} \\
\midrule
1 & Steps 1--4 & $\OPT \le 2$ & $\mathbf{2}: 2\;(\times 4)$ \\
2 & Steps 5--8 & $\OPT \le 8$ & $\mathbf{8}: 6+2\;(\times 4)$ \\
3 & Steps 9--11 & $\OPT \le 24$ & $\mathbf{24}: 20+2\tms 2$;\; $\mathbf{22}: 20+2\;(\times 2)$;\; $\mathbf{24}: 4\tms 6$ \\
4 & Step 12 & $\OPT \le 30$ & $\mathbf{30}: 24+6$;\; $\mathbf{30}: 20+6+2\tms 2$;\; $\mathbf{28}: 20+6+2\;(\times 2)$ \\
5 & Steps 13--16 & $\OPT \le 129$ & $\mathbf{129}: 99+[30]\;(\times 2)$;\; $\mathbf{127}: 99+[28]\;(\times 2)$ --- Row~4 plus one~$99$ per machine \\
6 & Steps 17--19 & $\OPT \le 392$ & $\mathbf{392}: 276+24+3\tms 20+4\tms 6+4\tms 2$;\; $\mathbf{375}: 276+99$;\; $\mathbf{276}: 276$;\; $\mathbf{297}: 3\tms 99$ \\
7 & Step 20 & $\OPT \le 415$ & $\mathbf{415}: 316+99$;\; $\mathbf{415}: 276+99+2\tms 20$;\; $\mathbf{415}: 276+99+24+2\tms 6+2\tms 2$;\newline $\mathbf{411}: 276+99+20+2\tms 6+2\tms 2$ \\
\midrule
8 & Case 1, Steps 21--24 & $\OPT \le 1985$ & $\mathbf{1985}: 1570+[415]\;(\times 3)$;\; $\mathbf{1941}: 1530+[411]$ --- Row~7 plus $3\tms 1570$ and $1530$ \\
9 & Cases 2--4, Steps 21--24 & $\OPT \le 1945$ & $\mathbf{1945}: 1530+[415]\;(\times 3)$;\; $\mathbf{1935}: 1524+[411]$ --- Row~7 plus $3\tms 1530$ and $1524$ of Case~2, which dominate Cases~3--4 \\
\midrule
10 & All Cases, Steps 25--27 & $\OPT \le 5774$ & $\mathbf{5774}: 4244+1530$;\; $\mathbf{5348}: 4244+2\tms[552]$;\; $\mathbf{4796}: 4244+[552]$;\; $\mathbf{4710}: 3\tms 1570$ \\
11 & Case 1, Step 28 & $\OPT \le 6366$ & $\mathbf{6366}: 4836+1530$;\; $\mathbf{6366}: 4244+1570+[552]\;(\times 3)$ \\
12 & Case 1, Step 29 & $\OPT \le 8488$ & $\mathbf{8488}: 8488$;\; $\mathbf{8488}: 2\tms 4244$;\; $\mathbf{8488}: 4836+1570+1530+[552]$;\newline $\mathbf{8488}: 4244+2\tms 1570+2\tms[552]$ \\
13 & Case 2, Step 28 & $\OPT \le 6246$ & $\mathbf{6246}: 4722+1524$;\; $\mathbf{6246}: 4164+1530+[552]\;(\times 3)$ \\
14 & Case 2, Step 29 & $\OPT \le 8328$ & $\mathbf{8328}: 8328$;\; $\mathbf{8328}: 2\tms 4164$;\; $\mathbf{8328}: 4722+1530+1524+[552]$;\newline $\mathbf{8328}: 4164+2\tms 1530+2\tms[552]$ \\
15 & Case 3, Step 28 & $\OPT \le 6228$ & $\mathbf{6228}: 4710+1512+6$;\; $\mathbf{6228}: 4152+1530+[546]$;\; $\mathbf{6228}: 4152+1524+[552]\;(\times 2)$ \\
16 & Case 3, Step 29 & $\OPT \le 8304$ & $\mathbf{8304}: 8304$;\; $\mathbf{8304}: 2\tms 4152$;\; $\mathbf{8304}: 4710+1530+1512+[552]$;\newline $\mathbf{8304}: 4152+2\tms 1524+2\tms[552]$ \\
17 & Case 4, Step 28 & $\OPT \le 6189$ & $\mathbf{6186}: 4680+1472+20+2\tms 6+2$;\; $\mathbf{6187}: 4125+1530+[532]$;\; $\mathbf{6187}: 4125+1524+[538]$;\newline $\mathbf{6189}: 4125+1512+[552]$ \\
18 & Case 4, Step 29 & $\OPT \le 8250$ & $\mathbf{8250}: 8250$;\; $\mathbf{8250}: 2\tms 4125$;\; $\mathbf{8250}: 4680+1530+1472+[568]$;\newline $\mathbf{8249}: 4125+1524+1512+[536]+[552]$ \\
\bottomrule
\end{tabular*}
\end{table}

%% file: tikz/fig_simple_branching.tex
\begin{figure}[!tb]
\centering
\begin{tikzpicture}[
  x=1cm, y=1cm,
  job/.style={draw=black, semithick},
  cut/.style={draw=black!70, densely dashed, semithick},
  bound/.style={dashed, line width=1.2pt},
  lbl/.style={font=\scriptsize, inner sep=0.5pt, align=center},
  toplbl/.style={font=\scriptsize, anchor=south, inner sep=1.5pt},
  subcap/.style={font=\scriptsize, anchor=north, inner sep=1.5pt},
  casetitle/.style={font=\footnotesize, anchor=north west, inner sep=0pt},
  frame/.style={draw=black, thick},
]
  \def\w{0.78}
  \def\s{4.16}
  \def\k{0.00088}%
  \def\CaseFrame#1#2#3#4#5{%
    \draw[frame] (-0.22,-0.82) rectangle ({\s+4*\w+0.22},2.42);
    \node[casetitle] at (-0.10,2.34) {#1};
    \pgfmathsetmacro{\yL}{#2*\k}%
    \node[lbl, anchor=north] at ({0.5*\w},-0.10) {$H$};
    \foreach \mi in {1,2,3} {\node[lbl, anchor=north] at ({(\mi+0.5)*\w},-0.10) {light};}
    \node[subcap] at ({2*\w},-0.40) {$\ALG$: $L = #2$};
    \node[font=\scriptsize] at ({0.5*(4*\w+\s)},0.52) {versus};
    \node[subcap] at ({\s+2*\w},-0.40) {$\OPT \le #4$ (Row~#5)};
  }
  \def\CaseLines{%
    \draw[bound] (-0.06,\yL) -- ({4*\w+0.06},\yL);
    \node[lbl, anchor=west] at ({4*\w+0.08},\yL) {$L$};
    \drawMachineTray{0}{0}{\w}
    \drawMachineTray{\s}{0}{\w}
  }
  \def\ACol#1#2#3#4#5#6{%
    \pgfmathsetmacro{\hb}{#2*\k}%
    \pgfmathsetmacro{\htop}{(#2+#5)*\k}%
    \pgfmathsetmacro{\hL}{min(\htop,\yL)}%
    \filldraw[job, fill=jobLightGray] ({#1*\w},0) rectangle ({(#1+1)*\w},\hb);
    \node[lbl] at ({(#1+0.5)*\w},{0.5*\hb}) {$#2$};
    \filldraw[job, fill=#3] ({#1*\w},\hb) rectangle ({(#1+1)*\w},\hL);
    \node[lbl, text=#6] at ({(#1+0.5)*\w},{0.5*(\hb+\hL)}) {#4};
    \ifdim\htop pt>\hL pt
      \fill[#3!35] ({#1*\w},\hL) rectangle ({(#1+1)*\w},\htop);
      \draw[cut] ({#1*\w},\hL) -- ({#1*\w},\htop) -- ({(#1+1)*\w},\htop) -- ({(#1+1)*\w},\hL);
    \fi
    \pgfmathtruncatemacro{\tot}{#2+#5}%
    \node[toplbl] at ({(#1+0.5)*\w},\htop) {$\tot$};
  }
  \def\OCol#1#2#3#4#5{%
    \pgfmathsetmacro{\hb}{#2*\k}%
    \pgfmathsetmacro{\htop}{(#2+#4)*\k}%
    \filldraw[job, fill=jobLightGray] ({\s+#1*\w},0) rectangle ({\s+(#1+1)*\w},\hb);
    \node[lbl] at ({\s+(#1+0.5)*\w},{0.5*\hb}) {$#2$};
    \filldraw[job, fill=#3] ({\s+#1*\w},\hb) rectangle ({\s+(#1+1)*\w},\htop);
    \node[lbl, text=#5] at ({\s+(#1+0.5)*\w},{0.5*(\hb+\htop)}) {$#4$};
    \pgfmathtruncatemacro{\tot}{#2+#4}%
    \node[toplbl] at ({\s+(#1+0.5)*\w},\htop) {$\tot$};
  }

  \begin{scope}[shift={(0,3.48)}]
    \CaseFrame{\textbf{Case 1:} $J_{21}$ on $H$}{1973}{1973}{1985}{8}
    \ACol{0}{443}{jobOrange}{$J_{21}$\\$1530$}{1530}{white}
    \ACol{1}{403}{white}{$J_{22}$\\$1570$}{1570}{black}
    \ACol{2}{403}{white}{$J_{23}$\\$1570$}{1570}{black}
    \ACol{3}{403}{white}{$J_{24}$\\$1570$}{1570}{black}
    \OCol{0}{415}{white}{1570}{black}
    \OCol{1}{415}{white}{1570}{black}
    \OCol{2}{415}{white}{1570}{black}
    \OCol{3}{411}{jobOrange}{1530}{white}
    \CaseLines
  \end{scope}

  \begin{scope}[shift={(7.96,3.48)}]
    \CaseFrame{\textbf{Case 2:} $J_{22}$ on $H$}{1933}{1967}{1945}{9}
    \ACol{0}{443}{jobOrange}{$J_{22}$\\$1524$}{1524}{white}
    \ACol{1}{403}{jobLightBlue}{$J_{21}$\\$1530$}{1530}{black}
    \ACol{2}{403}{white}{$J_{23}$\\$1530$}{1530}{black}
    \ACol{3}{403}{white}{$J_{24}$\\$1530$}{1530}{black}
    \OCol{0}{415}{jobLightBlue}{1530}{black}
    \OCol{1}{415}{white}{1530}{black}
    \OCol{2}{415}{white}{1530}{black}
    \OCol{3}{411}{jobOrange}{1524}{white}
    \CaseLines
  \end{scope}

  \begin{scope}[shift={(0,0)}]
    \CaseFrame{\textbf{Case 3:} $J_{23}$ on $H$}{1927}{1955}{1945}{9}
    \ACol{0}{443}{jobOrange}{$J_{23}$\\$1512$}{1512}{white}
    \ACol{1}{403}{jobLightBlue}{$J_{21}$\\$1530$}{1530}{black}
    \ACol{2}{403}{jobLightBlue}{$J_{22}$\\$1524$}{1524}{black}
    \ACol{3}{403}{white}{$J_{24}$\\$1524$}{1524}{black}
    \OCol{0}{415}{jobLightBlue}{1530}{black}
    \OCol{1}{415}{jobLightBlue}{1524}{black}
    \OCol{2}{415}{white}{1524}{black}
    \OCol{3}{411}{jobOrange}{1512}{white}
    \CaseLines
  \end{scope}

  \begin{scope}[shift={(7.96,0)}]
    \CaseFrame{\textbf{Case 4:} $J_{24}$ on $H$}{1915}{1933}{1945}{9}
    \ACol{0}{443}{jobOrange}{$J_{24}$\\$1472$}{1472}{white}
    \ACol{1}{403}{jobLightBlue}{$J_{21}$\\$1530$}{1530}{black}
    \ACol{2}{403}{jobLightBlue}{$J_{22}$\\$1524$}{1524}{black}
    \ACol{3}{403}{jobLightBlue}{$J_{23}$\\$1512$}{1512}{black}
    \OCol{0}{415}{jobLightBlue}{1530}{black}
    \OCol{1}{415}{jobLightBlue}{1524}{black}
    \OCol{2}{415}{jobLightBlue}{1512}{black}
    \OCol{3}{411}{jobOrange}{1472}{white}
    \CaseLines
  \end{scope}
\end{tikzpicture}
\caption{Phase~2 (layer~$5$, Steps~21--24) in Cases~1--4, starting from the state after Step~20 shown in \Cref{fig:easy_step20}. Left in each panel: the schedule of $\ALG$. Gray is the load after Phase~1. Orange is the job that $\ALG$ places on the heavy machine~$H$; it determines the case. Light blue are the jobs of layer~$5$ that arrived before it (probes); $\ALG$ placed them on light machines. White are the jobs with which the adversary completes the layer. The adversary then reduces all loads to their minimum~$L$ (dashed line), cutting off the dashed parts above~$L$. Right: an offline schedule of the same four layer-$5$ jobs (with matching colors) on top of the schedule $(415, 415, 415, 411)$ of \Cref{fig:easy_step20}(b), witnessing $\OPT \le 1985$ in Case~1 and $\OPT \le 1945$ in Cases~2--4 (Rows~8 and~9 of \Cref{tab:easy_opt}).}
\label{fig:easy_phase2}
\end{figure}
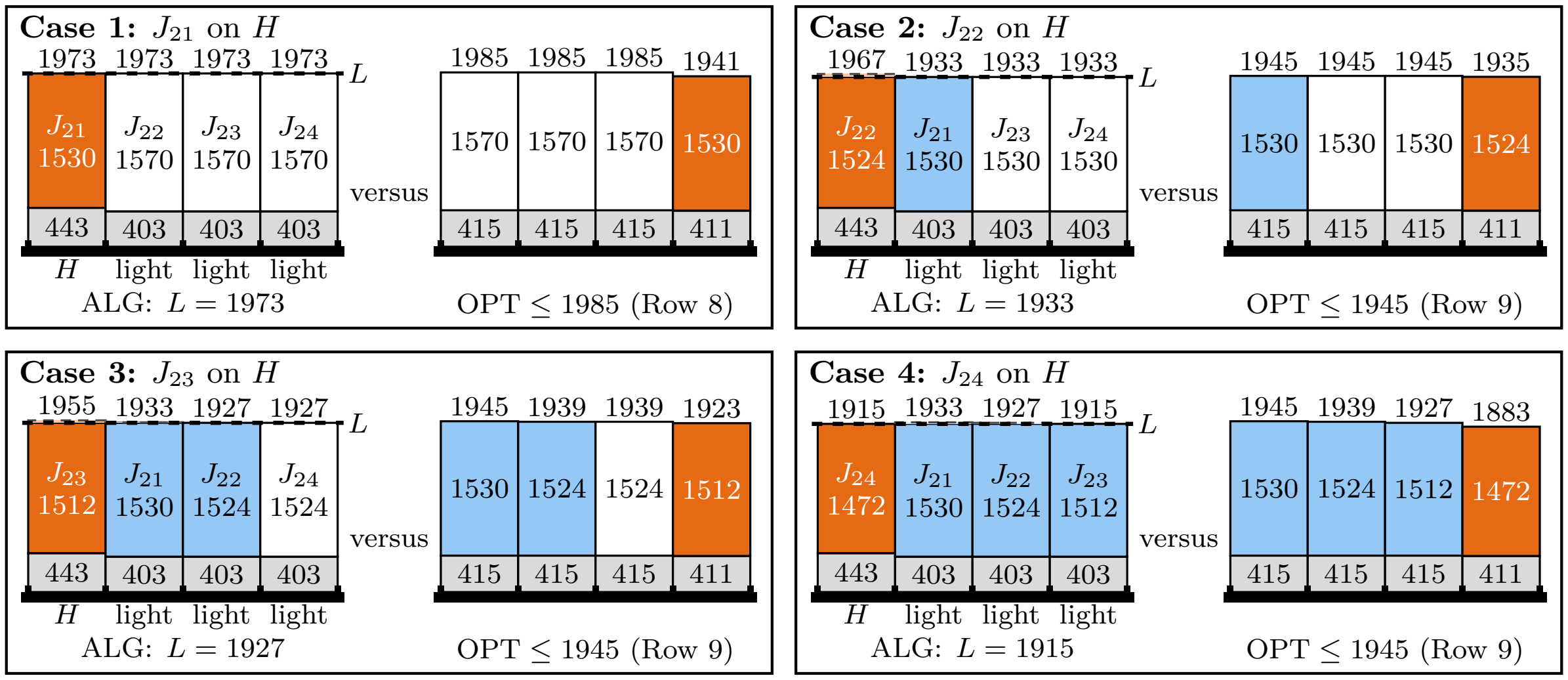

%% file: tikz/fig_simple_endgame.tex
\begin{figure}[!tb]
\centering
\begin{tikzpicture}[
  x=1cm, y=0.92cm,
  job/.style={draw=black, semithick},
  cut/.style={draw=black!70, densely dashed, semithick},
  bound/.style={dashed, line width=1.2pt},
  blbl/.style={font=\fontsize{5.2}{6}\selectfont, inner sep=0pt},
  lbl/.style={font=\fontsize{6}{7}\selectfont, inner sep=0pt},
  wlbl/.style={lbl, text=white},
  mlbl/.style={font=\scriptsize, inner sep=0pt},
  mwlbl/.style={mlbl, text=white},
  flbl/.style={font=\scriptsize, text=white, inner sep=0pt},
  toplbl/.style={font=\scriptsize, anchor=south, inner sep=1.5pt},
  subcap/.style={font=\scriptsize, anchor=north, inner sep=1.5pt},
  casetitle/.style={font=\footnotesize, anchor=north west, inner sep=0pt},
  frame/.style={draw=black, thick},
]
  \def\w{0.83}
  \def\s{4.46}
  \pgfmathsetmacro{\k}{3.90/14705}%
  \def\hBmin{0.24}%
  \def\yTop{3.90}%
  \def\JobBox#1#2#3#4#5#6#7{%
    \filldraw[job, fill=#5] ({#2+#1*\w},#3) rectangle ({#2+(#1+1)*\w},#4);
    \node[#6] at ({#2+(#1+0.5)*\w},{0.5*(#3+#4)}) {$#7$};
  }
  \def\EFrame#1#2#3#4{%
    \pgfmathtruncatemacro{\LtA}{#2+3*#3}%
    \pgfmathtruncatemacro{\tA}{2*#3}%
    \draw[frame] (-0.24,-0.60) rectangle ({\s+4*\w+0.24},{\yTop+0.74});
    \node[casetitle] at (-0.12,{\yTop+0.66}) {#1};
    \node[font=\scriptsize] at ({0.5*(4*\w+\s)},0.62) {versus};
    \node[subcap] at ({2*\w},-0.14) {$\ALG \ge L + 3a = \LtA$};
    \node[subcap] at ({\s+2*\w},-0.14) {$\OPT \le 2a = \tA$ (Row~#4)};
  }
  \def\EAlg#1#2#3{%
    \pgfmathtruncatemacro{\tA}{2*#2}%
    \pgfmathtruncatemacro{\LtA}{#1+3*#2}%
    \pgfmathtruncatemacro{\LA}{#1+#2}%
    \pgfmathtruncatemacro{\LB}{#1+#3}%
    \pgfmathsetmacro{\hL}{#1*\k}%
    \pgfmathsetmacro{\yLa}{(#1+#2)*\k}%
    \pgfmathsetmacro{\yLb}{(#1+#3)*\k}%
    \pgfmathsetmacro{\yLt}{(#1+3*#2)*\k}%
    \foreach \c in {0,1,2,3} {\JobBox{\c}{0}{0}{\hL}{jobLightGray}{lbl}{#1}}
    \foreach \c in {0,1,2} {\JobBox{\c}{0}{\hL}{\yLa}{jobMedBlue}{mwlbl}{#2}}
    \JobBox{0}{0}{\yLa}{\yLt}{jobOrange}{flbl}{\tA}
    \JobBox{3}{0}{\hL}{\yLa}{jobLightBlue}{mlbl}{#3}
    \fill[jobLightBlue!35] ({3*\w},\yLa) rectangle ({4*\w},\yLb);
    \draw[cut] ({3*\w},\yLa) -- ({3*\w},\yLb) -- ({4*\w},\yLb) -- ({4*\w},\yLa);
    \draw[bound] (-0.06,\yLa) -- ({4*\w+0.06},\yLa);
    \node[lbl, anchor=west] at ({4*\w+0.08},\yLa) {$L+a$};
    \node[toplbl] at ({0.5*\w},\yLt) {$\LtA$};
    \foreach \c in {1,2} {\node[toplbl] at ({(\c+0.5)*\w},\yLa) {$\LA$};}
    \node[toplbl] at ({3.5*\w},\yLb) {$\LB$};
    \drawMachineTray{0}{0}{\w}
  }
  \def\EOpt#1#2#3#4#5#6#7{%
    \pgfmathtruncatemacro{\tA}{2*#1}%
    \ifnum#7=0
      \def\bB{552}\def\bA{552}\def\bst{jobLightGray}%
    \else
      \def\bB{568}\def\bA{536}\def\bst{jobDarkGray!70}%
    \fi
    \pgfmathtruncatemacro{\loadB}{#2+#3+#4+\bB}%
    \pgfmathtruncatemacro{\loadA}{#1+#5+#6+\bA+552}%
    \pgfmathsetmacro{\yOpt}{2*#1*\k}%
    \pgfmathsetmacro{\ha}{#1*\k}%
    \pgfmathsetmacro{\yBa}{\hBmin+#3*\k}%
    \pgfmathsetmacro{\yBb}{\yBa+#4*\k}%
    \pgfmathsetmacro{\yBt}{\loadB*\k}
    \pgfmathsetmacro{\yAa}{2*\hBmin+#5*\k}%
    \pgfmathsetmacro{\yAb}{\yAa+#6*\k}%
    \pgfmathsetmacro{\yAt}{\loadA*\k}
    \JobBox{0}{\s}{0}{\yOpt}{jobOrange}{flbl}{\tA}
    \JobBox{1}{\s}{0}{\ha}{jobMedBlue}{mwlbl}{#1}
    \JobBox{1}{\s}{\ha}{\yOpt}{jobMedBlue}{mwlbl}{#1}
    \JobBox{2}{\s}{0}{\hBmin}{\bst}{blbl}{\bB}
    \JobBox{2}{\s}{\hBmin}{\yBa}{jobPaleBlue}{lbl}{#3}
    \JobBox{2}{\s}{\yBa}{\yBb}{jobPaleBlue}{lbl}{#4}
    \JobBox{2}{\s}{\yBb}{\yBt}{jobLightBlue}{mlbl}{#2}
    \JobBox{3}{\s}{0}{\hBmin}{\bst}{blbl}{\bA}
    \JobBox{3}{\s}{\hBmin}{2*\hBmin}{jobLightGray}{blbl}{552}
    \JobBox{3}{\s}{2*\hBmin}{\yAa}{jobPaleBlue}{lbl}{#5}
    \JobBox{3}{\s}{\yAa}{\yAb}{jobPaleBlue}{lbl}{#6}
    \JobBox{3}{\s}{\yAb}{\yAt}{jobMedBlue}{mwlbl}{#1}
    \foreach \c in {0,1} {\node[toplbl] at ({\s+(\c+0.5)*\w},\yOpt) {$\tA$};}
    \node[toplbl] at ({\s+2.5*\w},\yBt) {$\loadB$};
    \node[toplbl] at ({\s+3.5*\w},\yAt) {$\loadA$};
    \drawMachineTray{\s}{0}{\w}
  }
  \pgfmathsetmacro{\rowsep}{\yTop+1.58}
  \begin{scope}[shift={(0,\rowsep)}]
    \EFrame{\textbf{Case 1:} $L = 1973$, $a = 4244$, $b = 4836$}{1973}{4244}{12}
    \EAlg{1973}{4244}{4836}
    \EOpt{4244}{4836}{1570}{1530}{1570}{1570}{0}
  \end{scope}
  \begin{scope}[shift={(8.42,\rowsep)}]
    \EFrame{\textbf{Case 2:} $L = 1933$, $a = 4164$, $b = 4722$}{1933}{4164}{14}
    \EAlg{1933}{4164}{4722}
    \EOpt{4164}{4722}{1530}{1524}{1530}{1530}{0}
  \end{scope}
  \begin{scope}[shift={(0,0)}]
    \EFrame{\textbf{Case 3:} $L = 1927$, $a = 4152$, $b = 4710$}{1927}{4152}{16}
    \EAlg{1927}{4152}{4710}
    \EOpt{4152}{4710}{1530}{1512}{1524}{1524}{0}
  \end{scope}
  \begin{scope}[shift={(8.42,0)}]
    \EFrame{\textbf{Case 4:} $L = 1915$, $a = 4125$, $b = 4680$}{1915}{4125}{18}
    \EAlg{1915}{4125}{4680}
    \EOpt{4125}{4680}{1530}{1472}{1524}{1512}{1}
  \end{scope}
\end{tikzpicture}
\caption{Step~29 of the linear endgame (Phase~3) in Cases~1--4 (orange is the final job $J_{29} = 2a$, dark blue is~$a$, light blue is~$b$, pale blue are the four jobs of layer~$5$, and gray is the load~$L$ after Phase~2 and the Phase-1 blocks). Left in each panel: $\ALG$ starts layer~$6$ from common load~$L$, places $a, a, a, b$ on distinct machines (with the excess $b - a$ above $L + a$ cut off at the dashed line), and receives $J_{29} = 2a$, reaching makespan $L + 3a$. Right: the offline schedules of makespan~$2a$ from Rows~12, 14, 16, and~18 of \Cref{tab:easy_opt} using the four layer-$5$ jobs and the three $552$-blocks of \Cref{fig:easy_step20}(c), with $16$ units shifted between two blocks in Case~4 to form $568$ and~$536$. Case~3 is tight: $L + 3a = 14383 = c_0 \cdot 8304$.}
\label{fig:easy_endgame}
\end{figure}
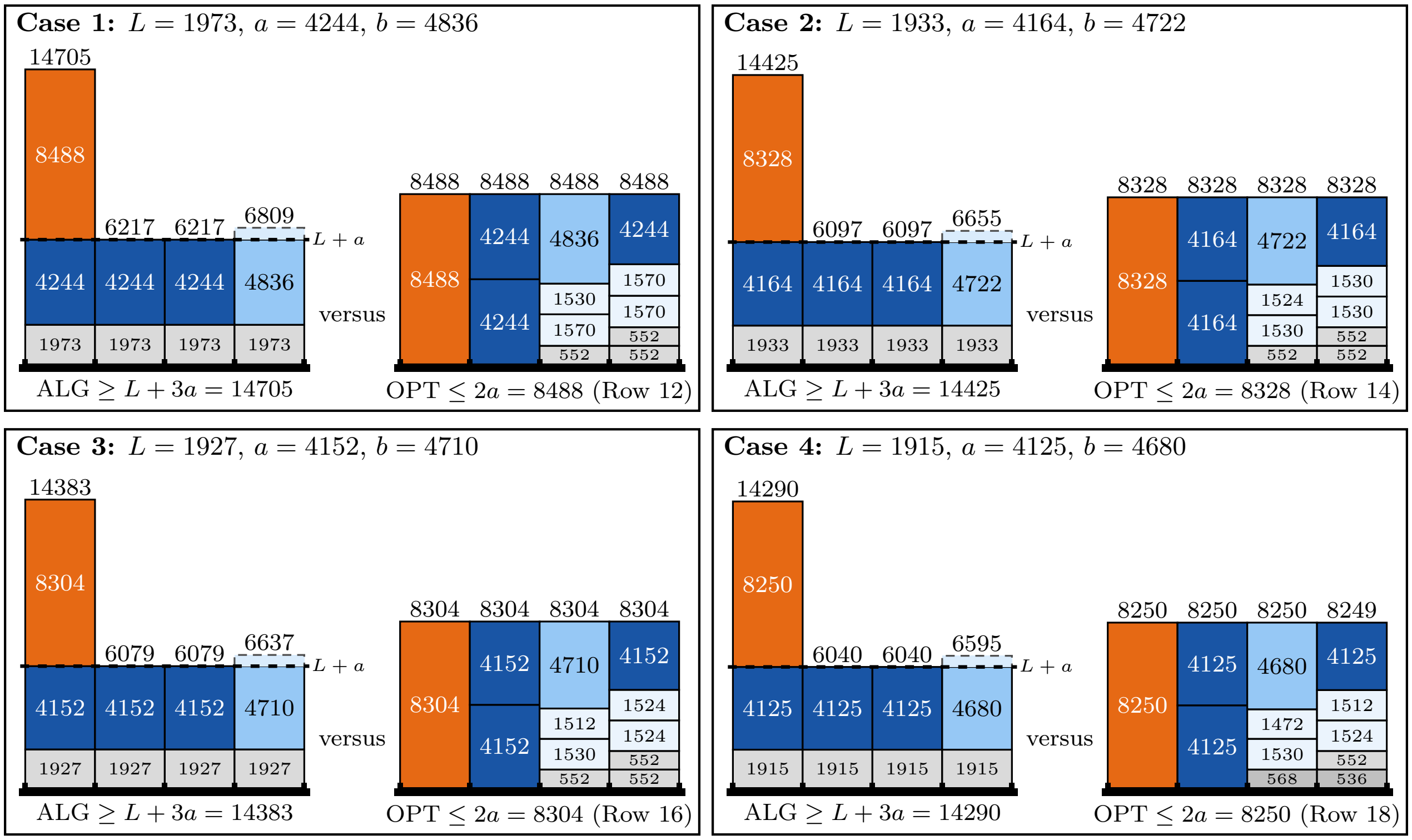

%% file: tables/tab07_opt_strategy.tex
\begin{table}[!tb]
\centering
\footnotesize
\setlength{\fboxsep}{1.1pt}
\renewcommand{\arraystretch}{1.12}
\setlength{\tabcolsep}{4pt}
\caption{The strategy of the adversary for \Cref{thm:opt}, where $c = \frac{14350}{8283}$ and $\OPT_t := \OPT(J_1, \dots, J_t)$. (a)~Phase~1: condition~\eqref{eq:linear_layer} for layers~$0$--$4$ using the minimum starting load~$\underline{l}_i$, with the upper bounds on $\OPT$ from \Cref{tab:phase1} in \Cref{sec:appendix_tables} (step in parentheses; in layers~$3$ and~$4$, marked~$^*$, one machine starts at $90$ or~$376$). (b)~Phase~2: in each case, $\ALG$ places the orange job on the heavy machine~$H$ and, before that, the decreasing probes (light blue) on light machines, where $^\dagger$ marks the probe placed on the $1176$-machine of state~$S_2$ (in Case~3a, marked~$^\ddagger$, either probe may use the $1176$-machine); the remaining jobs complete the layer. After reduction, the loads are $(h, h, h, l)$, or $L = 5898$ on all four machines in Case~1. (c)~Phase~3: in Case~1, a linear endgame $(a, a, a, b)$ with final job $f = 2a$. In Cases~2a--4d, the adversary presents the increasing probes $s_1 < s_2 < s_3 = s_1 + g$ until $\ALG$ places one on the light machine, which defines subcases~(i)--(iii), completes the layer with jobs of size $s_1 - g$ or $s_2 - g$ and the job~$b$, and ends with the final job~$f$. The loads before the final job are at least $E = l + s_1$, $l + s_2$, and $h + s_1$ in subcases (i), (ii), and~(iii); the tight branch, Case~3a(iii), determines~$c$. All immediate-loss conditions and offline schedules are listed step by step in \Cref{tab:phase2,tab:phase3a,tab:phase3b} of \Cref{sec:appendix_tables}.}
\label{tab:opt_strategy}

\begin{tabular*}{\linewidth}{@{\extracolsep{\fill}}cccccccc@{}}
\parthead{8}{\textbf{(a)}\enspace Phase~1: linear prefix without load reductions, layers $0$--$4$ (Steps 1--20)}
\toprule
Layer $i$ & Steps & $\underline{l}_i$ & $p_{4i+1}, \dots, p_{4i+4}$ & $j$ & $\OPT \le$ (Step) & $c \cdot \OPT$ & $\underline{l}_i + p + p_{4i+j}$ \\
\midrule
$0$ & 1--4   & $0$       & $4, 4, 4, 4$                  & $2$--$4$ & $4$\,(4)       & $6.93$    & $8$ \\
$1$ & 5--8   & $4$       & $19, 19, 19, 19$              & $2$--$4$ & $23$\,(8)      & $39.85$   & $42$ \\
$2$ & 9--12  & $23$      & $59, 59, 59, \mathbf{67}$     & $2, 3$   & $76$\,(11)     & $131.67$  & $141$ \\
    &        &           &                               & $4$      & $86$\,(12)     & $148.99$  & $149$ \\
$3$ & 13--16 & $82^{*}$  & $286, 286, 286, 286$          & $2$--$4$ & $372$\,(16)    & $644.48$  & $654$ \\
$4$ & 17--20 & $368^{*}$ & $800, 800, 800, \mathbf{918}$ & $2, 3$   & $1086$\,(19)   & $1881.46$ & $1968$ \\
    &        &           &                               & $4$      & $1204$\,(20)   & $2085.89$ & $2086$ \\
\bottomrule
\end{tabular*}

\vspace{5pt}
{\setlength{\tabcolsep}{3.5pt}%
\def\pdag#1{\probe{#1$^\dagger$}}%
\def\pddag#1{\probe{#1$^\ddagger$}}%
\begin{tabular*}{\linewidth}{@{\extracolsep{\fill}}ccccccccc@{}}
\parthead{9}{\textbf{(b)}\enspace Phase~2: branching layer~$5$ (Steps 21--24), starting from state $S_1$ or~$S_2$}
\toprule
 & & \multicolumn{4}{c}{Jobs of layer $5$} & & \multicolumn{2}{c}{Loads after reduction} \\
\cmidrule(lr){3-6}\cmidrule(l){8-9}
Case & State & $p_{21}$ & $p_{22}$ & $p_{23}$ & $p_{24}$ & $\OPT_{24} \le$ & heavier $h$ & light $l$ \\
\midrule
1  & $S_1, S_2$ & \onH{4620}    & \compl{4730}  & \compl{4730} & \compl{4730} & $5902$ & \multicolumn{2}{c}{all four: $L = 5898$} \\
2a & $S_1, S_2$ & \probe{4620}  & \onH{4578}    & \compl{4666} & \compl{4666} & $5834$ & $5834$ & $5788$ \\
2b & $S_2$      & \pdag{4620}   & \onH{4586}    & \compl{4674} & \compl{4674} & $5840$ & $5842$ & $5796$ \\
3a & $S_1, S_2$ & \pddag{4620}  & \pddag{4578}  & \onH{4535}   & \compl{4620} & $5790$ & $5788$ & $5746$ \\
3b & $S_2$      & \pdag{4620}  & \probe{4586} & \onH{4546}   & \compl{4628} & $5798$ & $5796$ & $5754$ \\
4a & $S_1$      & \probe{4620} & \probe{4578} & \probe{4535} & \onH{4452}   & $5747$ & $5746$ & $5703$ \\
4b & $S_2$      & \probe{4620} & \probe{4578} & \pdag{4535}  & \onH{4460}   & $5750$ & $5746$ & $5711$ \\
4c & $S_2$      & \probe{4620} & \pdag{4578}  & \probe{4535} & \onH{4468}   & $5751$ & $5754$ & $5703$ \\
4d & $S_2$      & \pdag{4620}  & \probe{4586} & \probe{4546} & \onH{4468}   & $5758$ & $5754$ & $5714$ \\
\bottomrule
\end{tabular*}}

\vspace{5pt}
{\setlength{\tabcolsep}{0pt}%
\fontsize{7.4}{9}\selectfont 
\thickmuskip=1.8mu\relax
\def\cs{{,}\kern0.5pt}%
\begin{tabular*}{\linewidth}{@{}l@{\hspace{5pt}}c@{\extracolsep{\fill}}cccccc@{}}
\parthead{8}{\textbf{(c)}\enspace Phase~3: layer~$6$ ($s_1, s_2, s_3$, $s_1 - g, s_2 - g$, $b$) and final job $J_{29} = f$ (Steps 25--29)}
\toprule
 & & \multicolumn{3}{c}{Jobs} & \multicolumn{3}{c}{Terminal bound $E + f \ge c \cdot \OPT_{29}$} \\
\cmidrule(lr){3-5}\cmidrule(l){6-8}
Case & $g$ & probes $s_1, s_2, s_3$ & $b$: (i), (ii), (iii) & $f$: (i), (ii), (iii) & (i) & (ii) & (iii) \\
\midrule
1  & ---  & \multicolumn{3}{c}{linear endgame: $a = 12652\,(\times 3),\; b = 14348,\; f = 25304$} & \multicolumn{3}{c}{$43854 > 43840.00$} \\
2a & $46$ & $12510\cs 12534\cs 12556$ & $14152\cs 14181\cs 14194$ & $24974\cs 25014\cs 25044$ & $43272 > 43266.56$ & $43336 > 43335.86$ & $43388 > 43387.83$ \\
2b & $46$ & $12529\cs 12550\cs 12575$ & $14170\cs 14196\cs 14210$ & $25007\cs 25045\cs 25079$ & $43332 > 43323.73$ & $43391 > 43389.56$ & $43450 > 43448.47$ \\
3a & $42$ & $12413\cs 12436\cs 12455$ & $14051\cs 14074\cs 14091$ & $24787\cs 24823\cs 24849$ & $42946 > 42942.59$ & $43005 > 43004.96$ & $\mathbf{43050 = 43050.00}$ \\
3b & $42$ & $12427\cs 12450\cs 12469$ & $14064\cs 14089\cs 14103$ & $24814\cs 24851\cs 24877$ & $42995 > 42989.36$ & $43055 > 43053.46$ & $43100 > 43098.51$ \\
4a & $43$ & $12321\cs 12345\cs 12364$ & $13951\cs 13968\cs 13982$ & $24604\cs 24640\cs 24666$ & $42628 > 42627.28$ & $42688 > 42687.92$ & $42733 > 42732.96$ \\
4b & $35$ & $12323\cs 12344\cs 12358$ & $13955\cs 13974\cs 13985$ & $24615\cs 24648\cs 24667$ & $42649 > 42648.07$ & $42703 > 42701.77$ & $42736 > 42734.69$ \\
4c & $51$ & $12330\cs 12359\cs 12381$ & $13954\cs 13979\cs 13998$ & $24614\cs 24659\cs 24689$ & $42647 > 42646.34$ & $42721 > 42720.83$ & $42773 > 42772.81$ \\
4d & $40$ & $12339\cs 12362\cs 12379$ & $13971\cs 13990\cs 14005$ & $24642\cs 24677\cs 24701$ & $42695 > 42694.84$ & $42753 > 42752.02$ & $42794 > 42793.60$ \\
\bottomrule
\end{tabular*}}
\end{table}

%% file: tables/tab02_opt_phase1.tex
{\scriptsize
\medmuskip=3mu\relax 
\relpenalty=10000\relax 
\setlength{\tabcolsep}{3.2pt}
\renewcommand{\arraystretch}{1.08}
\setlength{\LTcapwidth}{\textwidth}
\begin{longtable}{@{}r r >{\raggedright\arraybackslash}p{0.168\textwidth} >{\raggedright\arraybackslash}p{0.095\textwidth} r r r >{\raggedright\arraybackslash}p{0.400\textwidth}@{}}
\caption{\Cref{thm:opt}, Phase~1 (Steps 1--20): deterministic prefix of $20$ jobs ($c = \frac{14350}{8283}$), with immediate-stop bounds $\ALG \ge c \cdot \OPT$ and 4-machine offline schedules witnessing the upper bounds on $\OPT$ at every step. Machines without jobs are omitted.}\label{tab:phase1}\\
\toprule
Step & Job & Incoming $\ALG$ state(s) & Valid & $\ALG \ge$ & $\OPT \le$ & $c \cdot \OPT$ & Offline witness schedule ($\mathbf{load}:\text{jobs}$) \\
\midrule
\endfirsthead
\multicolumn{8}{@{}l}{\small\emph{\Cref{tab:phase1} (continued)}}\\[2pt]
\toprule
Step & Job & Incoming $\ALG$ state(s) & Valid & $\ALG \ge$ & $\OPT \le$ & $c \cdot \OPT$ & Offline witness schedule ($\mathbf{load}:\text{jobs}$) \\
\midrule
\endhead
\midrule
\multicolumn{8}{r@{}}{\emph{Continued on next page}}\\
\endfoot
\bottomrule
\endlastfoot
1 & $4$ & $(0, 0, 0, 0)$ & \Alt{any ($\le 0$)} & --- & $4$ & --- & \Mach{4}{4} \\
2 & $4$ & $(4, 0, 0, 0)$ & \Alt{$\le 0$} & $8$ & $4$ & $6.93$ & \Mach{4}{4\;(\times 2)} \\
3 & $4$ & $(4, 4, 0, 0)$ & \Alt{$\le 0$} & $8$ & $4$ & $6.93$ & \Mach{4}{4\;(\times 3)} \\
4 & $4$ & $(4, 4, 4, 0)$ & \Alt{$\le 0$} & $8$ & $4$ & $6.93$ & \Mach{4}{4\;(\times 4)} \\
\addlinespace[3pt]
5 & $19$ & $(4, 4, 4, 4)$ & \Alt{any ($\le 4$)} & --- & $19$ & --- & \Mach{19}{19}\Mach{8}{2\tms 4}\Mach{4}{4\;(\times 2)} \\
6 & $19$ & $(23, 4, 4, 4)$ & \Alt{$\le 4$} & $42$ & $19$ & $32.92$ & \Mach{19}{19\;(\times 2)}\Mach{8}{2\tms 4\;(\times 2)} \\
7 & $19$ & $(23, 23, 4, 4)$ & \Alt{$\le 4$} & $42$ & $19$ & $32.92$ & \Mach{19}{19\;(\times 3)}\Mach{16}{4\tms 4} \\
8 & $19$ & $(23, 23, 23, 4)$ & \Alt{$\le 4$} & $42$ & $23$ & $39.85$ & \Mach{23}{19 + 4\;(\times 4)} \\
\addlinespace[3pt]
9 & $59$ & $(23, 23, 23, 23)$ & \Alt{any ($\le 23$)} & --- & $59$ & --- & \Mach{59}{59}\Mach{38}{2\tms 19}\Mach{27}{19 + 2\tms 4\;(\times 2)} \\
10 & $59$ & $(82, 23, 23, 23)$ & \Alt{$\le 23$} & $141$ & $59$ & $102.22$ & \Mach{59}{59\;(\times 2)}\Mach{46}{2\tms 19 + 2\tms 4\;(\times 2)} \\
11 & $59$ & $(82, 82, 23, 23)$ & \Alt{$\le 23$} & $141$ & $76$ & $131.67$ & \Mach{76}{4\tms 19}\Mach{67}{59 + 2\tms 4}\Mach{63}{59 + 4\;(\times 2)} \\
12 & $67$ & $(82, 82, 82, 23)$ & \Alt{$\le 23$} & $149$ & $86$ & $148.99$ & \Mach{86}{67 + 19}\Mach{86}{59 + 19 + 2\tms 4}\Mach{82}{59 + 19 + 4\;(\times 2)} \\
\addlinespace[3pt]
13 & $286$ & $(90, 82, 82, 82)$ & \Alt{any ($\le 90$)} & --- & $286$ & --- & \Mach{286}{286}\Mach{118}{2\tms 59}\Mach{109}{67 + 2\tms 19 + 4}\Mach{109}{59 + 2\tms 19 + 3\tms 4} \\
14 & $286$ & $(376, 82, 82, 82)$\newline or $(368, 90, 82, 82)$ & \Alt{$\le 90$} & $654$ & $286$ & $495.48$ & \Mach{286}{286\;(\times 2)}\Mach{168}{67 + 59 + 2\tms 19 + 4}\Mach{168}{2\tms 59 + 2\tms 19 + 3\tms 4} \\
15 & $286$ & $(376, 368, 82, 82)$\newline or $(368, 368, 90, 82)$ & \Alt{$\le 90$} & $654$ & $305$ & $528.40$ & \Mach{305}{286 + 19}\Mach{301}{67 + 3\tms 59 + 3\tms 19}\Mach{294}{286 + 2\tms 4\;(\times 2)} \\
16 & $286$ & $(376, 368, 368, 82)$\newline or $(368, 368, 368, 90)$ & \Alt{$\le 90$} & $654$ & $372$ & $644.48$ & \Mach{372}{286 + 67 + 19}\Mach{372}{286 + 59 + 19 + 2\tms 4}\Mach{368}{286 + 59 + 19 + 4\;(\times 2)} \\
\addlinespace[3pt]
17 & $800$ & $(376, 368, 368, 368)$ & \Alt{any ($\le 376$)} & --- & $800$ & --- & \Mach{800}{800}\Mach{572}{2\tms 286}\Mach{454}{286 + 67 + 59 + 2\tms 19 + 4}\Mach{454}{286 + 2\tms 59 + 2\tms 19 + 3\tms 4} \\
18 & $800$ & $(1176, 368, 368, 368)$\newline or $(1168, 376, 368, 368)$ & \Alt{$\le 376$} & $1968$ & $800$ & $1385.97$ & \Mach{800}{800\;(\times 2)}\Mach{740}{2\tms 286 + 67 + 59 + 2\tms 19 + 4}\Mach{740}{2\tms 286 + 2\tms 59 + 2\tms 19 + 3\tms 4} \\
19 & $800$ & $(1176, 1168, 368, 368)$\newline or $(1168, 1168, 376, 368)$ & \Alt{$\le 376$} & $1968$ & $1086$ & $1881.46$ & \Mach{1086}{800 + 286}\Mach{936}{3\tms 286 + 59 + 19}\Mach{930}{800 + 2\tms 59 + 3\tms 4}\Mach{928}{800 + 67 + 3\tms 19 + 4} \\
20 & $918$ & $(1176, 1168, 1168, 368)$\newline or $(1168, 1168, 1168, 376)$ & \Alt{$\le 376$} & $2086$ & $1204$ & $2085.89$ & \Mach{1204}{918 + 286}\Mach{1204}{800 + 286 + 2\tms 59}\Mach{1195}{800 + 286 + 67 + 2\tms 19 + 4}\Mach{1195}{800 + 286 + 59 + 2\tms 19 + 3\tms 4} \\
\end{longtable}
}

%% file: tables/tab03_opt_phase2.tex
{\scriptsize
\medmuskip=3mu\relax 
\relpenalty=10000\relax 
\setlength{\tabcolsep}{3.2pt}
\renewcommand{\arraystretch}{1.08}
\setlength{\LTcapwidth}{\textwidth}
\begin{longtable}{@{}>{\raggedright\arraybackslash}p{0.10\textwidth} r >{\raggedright\arraybackslash}p{0.194\textwidth} r r r >{\raggedright\arraybackslash}p{0.395\textwidth}@{}}
\caption{\Cref{thm:opt}, Phase~2 (Steps 21--24 across all cases) and Phase~3 endgame for Case~1 (Steps 25--29): branching on which step places a job on the heavy machine ($\ge 1286$), with stop bounds $\ALG \ge c \cdot \OPT$ and offline schedules witnessing the upper bounds on $\OPT$.}\label{tab:phase2}\\
\toprule
Step / Case & Job & Valid / Transition & $\ALG \ge$ & $\OPT \le$ & $c \cdot \OPT$ & Offline witness schedule ($\mathbf{load}:\text{jobs}$) \\
\midrule
\endfirsthead
\multicolumn{7}{@{}l}{\small\emph{\Cref{tab:phase2} (continued)}}\\[2pt]
\toprule
Step / Case & Job & Valid / Transition & $\ALG \ge$ & $\OPT \le$ & $c \cdot \OPT$ & Offline witness schedule ($\mathbf{load}:\text{jobs}$) \\
\midrule
\endhead
\midrule
\multicolumn{7}{r@{}}{\emph{Continued on next page}}\\
\endfoot
\bottomrule
\endlastfoot
Step 21\newline(probe 1) & $4620$ & \Alt{$1168 \to$ Step 22a}\Alt{$1176 \to$ Step 22b}\Alt{$1286, 1294 \to$ Case 1} & --- & $4620$ & --- & \Mach{4620}{4620}\Mach{1600}{2\tms 800}\Mach{1599}{918 + 2\tms 286 + 59 + 2\tms 19 + 3\tms 4}\Mach{1599}{800 + 2\tms 286 + 67 + 2\tms 59 + \mbox{$2\tms 19 + 4$}} \\
Case 1,\newline Step 22 & $4730$ & \Alt{$\le 1176$} & $10636$ & $4730$ & $8194.56$ & \Mach{4730}{4730}\Mach{4620}{4620}\Mach{2399}{918 + 800 + 2\tms 286 + 59 + \mbox{$2\tms 19 + 3\tms 4$}}\Mach{2399}{2\tms 800 + 2\tms 286 + 67 + 2\tms 59 + \mbox{$2\tms 19 + 4$}} \\
Case 1,\newline Step 23 & $4730$ & \Alt{$\le 1176$} & $10628$ & $4730$ & $8194.56$ & \Mach{4730}{4730}\Mach{4730}{4730}\Mach{4710}{4620 + 59 + 19 + 3\tms 4}\Mach{4708}{918 + 3\tms 800 + 4\tms 286 + 67 + 2\tms 59 + \mbox{$3\tms 19 + 4$}} \\
Case 1,\newline Step 24 & $4730$ & \Alt{$\le 1176 \to$ Case 1, Step 25 (end of table), \mbox{$(5914, 5898, 5898, 5898)$} or \mbox{$(5906, 5906, 5898, 5898)$}} & $10628$ & $5902$ & $10225.00$ & \Mach{5902}{4730 + 800 + 286 + 67 + 19}\Mach{5902}{4730 + 800 + 286 + 59 + \mbox{$19 + 2\tms 4$}}\Mach{5902}{4730 + 800 + 286 + 59 + \mbox{$19 + 2\tms 4$}}\Mach{5902}{4620 + 918 + 286 + 59 + 19} \\
\addlinespace[5pt]
Step 22a\newline($4620$ on $1168$) & $4578$ & \Alt{$1168 \to$ Step 23a}\Alt{$1176 \to$ Step 23b}\Alt{$1286, 1294 \to$ Case 2a} & $10366$ & $4620$ & $8003.98$ & \Mach{4620}{4620}\Mach{4578}{4578}\Mach{2399}{918 + 800 + 2\tms 286 + 59 + \mbox{$2\tms 19 + 3\tms 4$}}\Mach{2399}{2\tms 800 + 2\tms 286 + 67 + 2\tms 59 + \mbox{$2\tms 19 + 4$}} \\
Case 2a,\newline Step 23 & $4666$ & \Alt{$\le 1176$} & $10454$ & $4666$ & $8083.68$ & \Mach{4666}{4666}\Mach{4666}{4620 + 2\tms 19 + 2\tms 4}\Mach{4666}{918 + 3\tms 800 + 4\tms 286 + 67 + \mbox{$2\tms 59 + 19$}}\Mach{4664}{4578 + 59 + 19 + 2\tms 4} \\
Case 2a,\newline Step 24 & $4666$ & \Alt{$\le 1176 \to$ Case 2a (\Cref{tab:phase3a}), \mbox{$(5872, 5834, 5834, 5788)$} or \mbox{$(5864, 5842, 5834, 5788)$}} & $10454$ & $5834$ & $10107.20$ & \Mach{5834}{4666 + 800 + 286 + 59 + \mbox{$19 + 4$}}\Mach{5832}{4620 + 800 + 286 + 67 + 59}\Mach{5832}{4578 + 918 + 286 + 2\tms 19 + 3\tms 4}\Mach{5830}{4666 + 800 + 286 + 59 + 19} \\
\addlinespace[5pt]
Step 22b\newline($4620$ on $1176$) & $4586$ & \Alt{$1168 \to$ Step 23c}\Alt{$1286 \to$ Case 2b} & $10382$ & $4620$ & $8003.98$ & \Mach{4620}{4620}\Mach{4586}{4586}\Mach{2399}{918 + 800 + 2\tms 286 + 59 + \mbox{$2\tms 19 + 3\tms 4$}}\Mach{2399}{2\tms 800 + 2\tms 286 + 67 + 2\tms 59 + \mbox{$2\tms 19 + 4$}} \\
Case 2b,\newline Step 23 & $4674$ & \Alt{$\le 1168$} & $10470$ & $4674$ & $8097.54$ & \Mach{4674}{4674}\Mach{4670}{4620 + 2\tms 19 + 3\tms 4}\Mach{4668}{4586 + 59 + 19 + 4}\Mach{4666}{918 + 3\tms 800 + 4\tms 286 + 67 + \mbox{$2\tms 59 + 19$}} \\
Case 2b,\newline Step 24 & $4674$ & \Alt{$\le 1168 \to$ Case 2b (\Cref{tab:phase3a}), \mbox{$(5872, 5842, 5842, 5796)$}} & $10470$ & $5840$ & $10117.59$ & \Mach{5840}{4620 + 800 + 286 + 67 + \mbox{$59 + 2\tms 4$}}\Mach{5838}{4674 + 800 + 286 + 59 + 19}\Mach{5838}{4674 + 800 + 286 + 59 + 19}\Mach{5836}{4586 + 918 + 286 + 2\tms 19 + 2\tms 4} \\
\addlinespace[5pt]
Step 23a\newline($4578$ on $1168$) & $4535$ & \Alt{$1168 \to$ Case 4a}\Alt{$1176 \to$ Case 4b}\Alt{$1286, 1294 \to$ Case 3a} & $10281$ & $4637$ & $8033.44$ & \Mach{4637}{4578 + 59}\Mach{4632}{4620 + 3\tms 4}\Mach{4632}{4535 + 59 + 2\tms 19}\Mach{4630}{918 + 3\tms 800 + 4\tms 286 + 67 + 59 + \mbox{$2\tms 19 + 4$}} \\
Step 23b\newline($4578$ on $1176$) & $4535$ & \Alt{$1168 \to$ Case 4c}\Alt{$1286 \to$ Case 3a} & $10289$ & $4637$ & $8033.44$ & \Mach{4637}{4578 + 59}\Mach{4632}{4620 + 3\tms 4}\Mach{4632}{4535 + 59 + 2\tms 19}\Mach{4630}{918 + 3\tms 800 + 4\tms 286 + 67 + 59 + \mbox{$2\tms 19 + 4$}} \\
Case 3a,\newline Step 24 & $4620$ & \Alt{$\le 1176 \to$ Case 3a (\Cref{tab:phase3a}), \mbox{$(5829, 5788, 5788, 5746)$}, \mbox{$(5821, 5796, 5788, 5746)$}, or \mbox{$(5821, 5788, 5788, 5754)$}} & $10366$ & $5790$ & $10030.97$ & \Mach{5790}{4578 + 800 + 286 + 67 + 59}\Mach{5788}{4620 + 800 + 286 + 59 + \mbox{$19 + 4$}}\Mach{5788}{4620 + 800 + 286 + 59 + \mbox{$19 + 4$}}\Mach{5785}{4535 + 918 + 286 + 2\tms 19 + 2\tms 4} \\
\addlinespace[5pt]
Step 23c\newline($4586$ on $1168$) & $4546$ & \Alt{$1168 \to$ Case 4d}\Alt{$1286 \to$ Case 3b} & $10300$ & $4645$ & $8047.30$ & \Mach{4645}{4586 + 59}\Mach{4639}{4620 + 19}\Mach{4634}{918 + 3\tms 800 + 4\tms 286 + 67 + 59 + \mbox{$2\tms 19 + 2\tms 4$}}\Mach{4632}{4546 + 59 + 19 + 2\tms 4} \\
Case 3b,\newline Step 24 & $4628$ & \Alt{$\le 1168 \to$ Case 3b (\Cref{tab:phase3a}), \mbox{$(5832, 5796, 5796, 5754)$}} & $10382$ & $5798$ & $10044.83$ & \Mach{5798}{4586 + 800 + 286 + 67 + 59}\Mach{5796}{4628 + 800 + 286 + 59 + \mbox{$19 + 4$}}\Mach{5792}{4620 + 800 + 286 + 59 + \mbox{$19 + 2\tms 4$}}\Mach{5792}{4546 + 918 + 286 + 2\tms 19 + 4} \\
\addlinespace[5pt]
Case 4a,\newline Step 24 & $4452$ & \Alt{$1294 \to$ Case 4a (\Cref{tab:phase3b}), \mbox{$(5788, 5746, 5746, 5703)$}} & $10155$ & $5747$ & $9956.47$ & \Mach{5747}{4535 + 800 + 286 + 67 + 59}\Mach{5746}{4578 + 800 + 286 + 59 + \mbox{$19 + 4$}}\Mach{5746}{4452 + 918 + 286 + 59 + \mbox{$19 + 3\tms 4$}}\Mach{5744}{4620 + 800 + 286 + 2\tms 19} \\
Case 4b,\newline Step 24 & $4460$ & \Alt{$1286 \to$ Case 4b (\Cref{tab:phase3b}), \mbox{$(5788, 5746, 5746, 5711)$}} & $10171$ & $5750$ & $9961.67$ & \Mach{5750}{4578 + 800 + 286 + 59 + \mbox{$19 + 2\tms 4$}}\Mach{5748}{4620 + 800 + 286 + 2\tms 19 + 4}\Mach{5747}{4535 + 800 + 286 + 67 + 59}\Mach{5746}{4460 + 918 + 286 + 59 + \mbox{$19 + 4$}} \\
Case 4c,\newline Step 24 & $4468$ & \Alt{$1286 \to$ Case 4c (\Cref{tab:phase3b}), \mbox{$(5788, 5754, 5754, 5703)$}} & $10171$ & $5751$ & $9963.40$ & \Mach{5751}{4535 + 800 + 286 + 67 + \mbox{$59 + 4$}}\Mach{5750}{4578 + 800 + 286 + 59 + \mbox{$19 + 2\tms 4$}}\Mach{5750}{4468 + 918 + 286 + 59 + 19}\Mach{5748}{4620 + 800 + 286 + 2\tms 19 + 4} \\
Case 4d,\newline Step 24 & $4468$ & \Alt{$1286 \to$ Case 4d (\Cref{tab:phase3b}), \mbox{$(5796, 5754, 5754, 5714)$}} & $10182$ & $5758$ & $9975.53$ & \Mach{5758}{4546 + 800 + 286 + 67 + 59}\Mach{5754}{4586 + 800 + 286 + 59 + \mbox{$19 + 4$}}\Mach{5754}{4468 + 918 + 286 + 59 + \mbox{$19 + 4$}}\Mach{5752}{4620 + 800 + 286 + 2\tms 19 + 2\tms 4} \\
\midrule
\multicolumn{7}{@{}l}{\emph{Phase~3 linear endgame for Case~1 (Steps 25--29)}}\\
\midrule
Case 1,\newline Step 25 & $12652$ & \Alt{any ($L = 5898$)} & --- & $12652$ & --- & \Mach{12652}{12652}\Mach{9350}{4730 + 4620}\Mach{7129}{4730 + 918 + 800 + 2\tms 286 + 59 + \mbox{$2\tms 19 + 3\tms 4$}}\Mach{7129}{4730 + 2\tms 800 + 2\tms 286 + 67 + 2\tms 59 + \mbox{$2\tms 19 + 4$}} \\
Case 1,\newline Step 26 & $12652$ & \Alt{$L = 5898$} & $31202$ & $12652$ & $21919.14$ & \Mach{12652}{12652}\Mach{12652}{12652}\Mach{11804}{2\tms 4730 + 2\tms 800 + 2\tms 286 + 67 + 59 + \mbox{$2\tms 19 + 2\tms 4$}}\Mach{11804}{4730 + 4620 + 918 + 800 + 2\tms 286 + 2\tms 59 + \mbox{$2\tms 19 + 2\tms 4$}} \\
Case 1,\newline Step 27 & $12652$ & \Alt{$L = 5898$} & $31202$ & $17272$ & $29923.12$ & \Mach{17272}{12652 + 4620}\Mach{14765}{12652 + 918 + 800 + 286 + 59 + \mbox{$2\tms 19 + 3\tms 4$}}\Mach{14765}{12652 + 2\tms 800 + 286 + 67 + 2\tms 59 + \mbox{$2\tms 19 + 4$}}\Mach{14762}{3\tms 4730 + 2\tms 286} \\
Case 1,\newline Step 28 & $14348$ & \Alt{$L = 5898$} & $32898$ & $18982$ & $32885.63$ & \Mach{18982}{12652 + 4730 + 2\tms 800}\Mach{18977}{12652 + 4730 + 918 + 2\tms 286 + 59 + \mbox{$2\tms 19 + 2\tms 4$}}\Mach{18977}{12652 + 4730 + 800 + 2\tms 286 + 67 + \mbox{$2\tms 59 + 2\tms 19$}}\Mach{18976}{14348 + 4620 + 2\tms 4} \\
Case 1,\newline Step 29 & $25304$ & \Alt{\emph{(terminal stop)}} & $43854$ & $25305$ & $43840.00$ & \Mach{25305}{14348 + 4730 + 4620 + 800 + 2\tms 286 + 67 + 2\tms 59 + \mbox{$2\tms 19 + 3\tms 4$}}\Mach{25304}{25304}\Mach{25304}{2\tms 12652}\Mach{25303}{12652 + 2\tms 4730 + 918 + 2\tms 800 + 2\tms 286 + 59 + \mbox{$2\tms 19 + 4$}} \\
\end{longtable}
}

%% file: tables/tab04_opt_phase3a.tex
{\scriptsize
\medmuskip=3mu\relax 
\relpenalty=10000\relax 
\setlength{\tabcolsep}{3.1pt}
\renewcommand{\arraystretch}{1.08}
\setlength{\LTcapwidth}{\textwidth}
\begin{longtable}{@{}>{\raggedright\arraybackslash}p{0.115\textwidth} r >{\raggedright\arraybackslash}p{0.172\textwidth} r r r >{\raggedright\arraybackslash}p{0.395\textwidth}@{}}
\caption{\Cref{thm:opt}, Phase~3 endgame (Steps 25--29) for Cases 2a, 2b, 3a, and 3b across Subcases (i), (ii), and (iii), starting from the reduced Phase-$2$ loads $(h, h, h, l)$ of \Cref{tab:opt_strategy}(b). Case~3a(iii) at Step~29 attains the tight bottleneck ratio $43050/24849 = 14350/8283 = c$.}\label{tab:phase3a}\\
\toprule
Case, Step & Job & Valid / Transition & $\ALG \ge$ & $\OPT \le$ & $c \cdot \OPT$ & Offline witness schedule ($\mathbf{load}:\text{jobs}$) \\
\midrule
\endfirsthead
\multicolumn{7}{@{}l}{\small\emph{\Cref{tab:phase3a} (continued)}}\\[2pt]
\toprule
Case, Step & Job & Valid / Transition & $\ALG \ge$ & $\OPT \le$ & $c \cdot \OPT$ & Offline witness schedule ($\mathbf{load}:\text{jobs}$) \\
\midrule
\endhead
\midrule
\multicolumn{7}{r@{}}{\emph{Continued on next page}}\\
\endfoot
\bottomrule
\endlastfoot
2a, 25 & $12510$ & \Alt{$5788 \to$ (i)}\Alt{$5834 \to$ (ii)--(iii)} & --- & $12510$ & --- & \Mach{12510}{12510}\Mach{9198}{4620 + 4578}\Mach{7065}{4666 + 918 + 800 + 2\tms 286 + 59 + \mbox{$2\tms 19 + 3\tms 4$}}\Mach{7065}{4666 + 2\tms 800 + 2\tms 286 + 67 + 2\tms 59 + \mbox{$2\tms 19 + 4$}} \\
2a(i), 26 & $12464$ & \Alt{$\le 5834$} & $30762$ & $12510$ & $21673.13$ & \Mach{12510}{12510}\Mach{12464}{12464}\Mach{11666}{4666 + 4620 + 2\tms 800 + 2\tms 286 + 67 + 2\tms 59 + \mbox{$19 + 4$}}\Mach{11662}{4666 + 4578 + 918 + 800 + 2\tms 286 + 59 + \mbox{$3\tms 19 + 3\tms 4$}} \\
2a(i), 27 & $12464$ & \Alt{$\le 5834$} & $30762$ & $17042$ & $29524.65$ & \Mach{17042}{12464 + 4578}\Mach{14581}{12510 + 2\tms 800 + 286 + 67 + 2\tms 59}\Mach{14573}{12464 + 918 + 800 + 286 + 59 + \mbox{$2\tms 19 + 2\tms 4$}}\Mach{14570}{2\tms 4666 + 4620 + 2\tms 286 + 2\tms 19 + 2\tms 4} \\
2a(i), 28 & $14152$ & \Alt{$\le 5834$} & $32450$ & $18730$ & $32449.05$ & \Mach{18730}{14152 + 4578}\Mach{18730}{12464 + 4666 + 2\tms 800}\Mach{18729}{12510 + 4620 + 918 + 2\tms 286 + 59 + \mbox{$2\tms 19 + 3\tms 4$}}\Mach{18729}{12464 + 4666 + 800 + 2\tms 286 + 67 + 2\tms 59 + \mbox{$2\tms 19 + 4$}} \\
2a(i), 29 & $24974$ & \Alt{\emph{(terminal stop)}} & $43272$ & $24974$ & $43266.56$ & \Mach{24974}{24974}\Mach{24974}{2\tms 12464 + 2\tms 19 + 2\tms 4}\Mach{24972}{14152 + 4666 + 4578 + 800 + 2\tms 286 + 67 + \mbox{$2\tms 59 + 19$}}\Mach{24972}{12510 + 4666 + 4620 + 918 + 2\tms 800 + 2\tms 286 + 59 + \mbox{$19 + 2\tms 4$}} \\
2a(ii)--(iii), 26 & $12534$ & \Alt{$5788 \to$ (ii)}\Alt{$5834 \to$ (iii)} & $30878$ & $12534$ & $21714.70$ & \Mach{12534}{12534}\Mach{12510}{12510}\Mach{11666}{4666 + 4620 + 2\tms 800 + 2\tms 286 + 67 + 2\tms 59 + \mbox{$19 + 4$}}\Mach{11662}{4666 + 4578 + 918 + 800 + 2\tms 286 + 59 + \mbox{$3\tms 19 + 3\tms 4$}} \\
2a(ii), 27 & $12488$ & \Alt{$\le 5834$} & $30810$ & $17066$ & $29566.23$ & \Mach{17066}{12488 + 4578}\Mach{14602}{2\tms 4666 + 4620 + 2\tms 286 + 59 + 19}\Mach{14596}{12534 + 2\tms 800 + 286 + 67 + 59 + \mbox{$2\tms 19 + 3\tms 4$}}\Mach{14596}{12510 + 918 + 800 + 286 + 59 + \mbox{$19 + 4$}} \\
2a(ii), 28 & $14181$ & \Alt{$\le 5834$} & $32503$ & $18761$ & $32502.76$ & \Mach{18761}{12534 + 4666 + 800 + 2\tms 286 + 67 + \mbox{$2\tms 59 + 4$}}\Mach{18761}{12510 + 4620 + 2\tms 800 + 19 + 3\tms 4}\Mach{18760}{12488 + 4666 + 918 + 2\tms 286 + \mbox{$59 + 3\tms 19$}}\Mach{18759}{14181 + 4578} \\
2a(ii), 29 & $25014$ & \Alt{\emph{(terminal stop)}} & $43336$ & $25014$ & $43335.86$ & \Mach{25014}{25014}\Mach{25014}{14181 + 4666 + 4620 + 800 + 2\tms 286 + \mbox{$2\tms 59 + 3\tms 19$}}\Mach{25014}{12510 + 12488 + 4\tms 4}\Mach{25013}{12534 + 4666 + 4578 + 918 + 2\tms 800 + 2\tms 286 + 67 + \mbox{$59 + 19$}} \\
2a(iii), 27 & $12556$ & \Alt{$\le 5834$} & $30900$ & $17088$ & $29604.35$ & \Mach{17088}{12510 + 4578}\Mach{14616}{12534 + 918 + 800 + 286 + \mbox{$59 + 19$}}\Mach{14614}{12556 + 2\tms 800 + 286 + 67 + 59 + \mbox{$2\tms 19 + 2\tms 4$}}\Mach{14610}{2\tms 4666 + 4620 + 2\tms 286 + 59 + \mbox{$19 + 2\tms 4$}} \\
2a(iii), 28 & $14194$ & \Alt{$\le 5834$} & $32538$ & $18781$ & $32537.41$ & \Mach{18781}{12556 + 4666 + 800 + 2\tms 286 + 67 + 59 + \mbox{$3\tms 19 + 4$}}\Mach{18781}{12534 + 4620 + 918 + 2\tms 286 + \mbox{$2\tms 59 + 19$}}\Mach{18780}{14194 + 4578 + 2\tms 4}\Mach{18780}{12510 + 4666 + 2\tms 800 + 4} \\
2a(iii), 29 & $25044$ & \Alt{\emph{(terminal stop)}} & $43388$ & $25044$ & $43387.83$ & \Mach{25044}{25044}\Mach{25044}{12534 + 12510}\Mach{25041}{14194 + 4666 + 4578 + 800 + 2\tms 286 + 67 + 2\tms 59 + \mbox{$2\tms 19 + 2\tms 4$}}\Mach{25037}{12556 + 4666 + 4620 + 918 + 2\tms 800 + 2\tms 286 + 59 + \mbox{$2\tms 19 + 2\tms 4$}} \\
\addlinespace[5pt]
2b, 25 & $12529$ & \Alt{$5796 \to$ (i)}\Alt{$5842 \to$ (ii)--(iii)} & --- & $12529$ & --- & \Mach{12529}{12529}\Mach{9206}{4620 + 4586}\Mach{7073}{4674 + 918 + 800 + 2\tms 286 + 59 + \mbox{$2\tms 19 + 3\tms 4$}}\Mach{7073}{4674 + 2\tms 800 + 2\tms 286 + 67 + 2\tms 59 + \mbox{$2\tms 19 + 4$}} \\
2b(i), 26 & $12483$ & \Alt{$\le 5842$} & $30808$ & $12529$ & $21706.04$ & \Mach{12529}{12529}\Mach{12483}{12483}\Mach{11678}{4674 + 4620 + 2\tms 800 + 2\tms 286 + 67 + 2\tms 59 + \mbox{$19 + 2\tms 4$}}\Mach{11674}{4674 + 4586 + 918 + 800 + 2\tms 286 + 59 + \mbox{$3\tms 19 + 2\tms 4$}} \\
2b(i), 27 & $12483$ & \Alt{$\le 5842$} & $30808$ & $17069$ & $29571.43$ & \Mach{17069}{12483 + 4586}\Mach{14599}{2\tms 4674 + 4620 + 2\tms 286 + 59}\Mach{14591}{12529 + 2\tms 800 + 286 + 67 + 59 + \mbox{$2\tms 19 + 3\tms 4$}}\Mach{14588}{12483 + 918 + 800 + 286 + 59 + \mbox{$2\tms 19 + 4$}} \\
2b(i), 28 & $14170$ & \Alt{$\le 5842$} & $32495$ & $18756$ & $32494.10$ & \Mach{18756}{14170 + 4586}\Mach{18756}{12483 + 4674 + 918 + 2\tms 286 + 59 + \mbox{$2\tms 19 + 3\tms 4$}}\Mach{18753}{12529 + 4620 + 2\tms 800 + 4}\Mach{18752}{12483 + 4674 + 800 + 2\tms 286 + 67 + \mbox{$2\tms 59 + 2\tms 19$}} \\
2b(i), 29 & $25007$ & \Alt{\emph{(terminal stop)}} & $43332$ & $25007$ & $43323.73$ & \Mach{25007}{25007}\Mach{25007}{12529 + 4674 + 4620 + 918 + 2\tms 800 + 2\tms 286 + 59 + \mbox{$19 + 4\tms 4$}}\Mach{25006}{14170 + 4674 + 4586 + 800 + 2\tms 286 + 67 + \mbox{$2\tms 59 + 19$}}\Mach{25004}{2\tms 12483 + 2\tms 19} \\
2b(ii)--(iii), 26 & $12550$ & \Alt{$5796 \to$ (ii)}\Alt{$5842 \to$ (iii)} & $30921$ & $12550$ & $21742.42$ & \Mach{12550}{12550}\Mach{12529}{12529}\Mach{11678}{4674 + 4620 + 2\tms 800 + 2\tms 286 + 67 + 2\tms 59 + \mbox{$19 + 2\tms 4$}}\Mach{11674}{4674 + 4586 + 918 + 800 + 2\tms 286 + 59 + \mbox{$3\tms 19 + 2\tms 4$}} \\
2b(ii), 27 & $12504$ & \Alt{$\le 5842$} & $30850$ & $17090$ & $29607.81$ & \Mach{17090}{12504 + 4586}\Mach{14618}{2\tms 4674 + 4620 + 2\tms 286 + 59 + 19}\Mach{14615}{12529 + 918 + 800 + 286 + 59 + \mbox{$19 + 4$}}\Mach{14612}{12550 + 2\tms 800 + 286 + 67 + 59 + \mbox{$2\tms 19 + 3\tms 4$}} \\
2b(ii), 28 & $14196$ & \Alt{$\le 5842$} & $32542$ & $18783$ & $32540.87$ & \Mach{18783}{12550 + 4674 + 800 + 2\tms 286 + 67 + 59 + \mbox{$3\tms 19 + 4$}}\Mach{18783}{12529 + 4674 + 918 + 2\tms 286 + 59 + \mbox{$19 + 3\tms 4$}}\Mach{18783}{12504 + 4620 + 2\tms 800 + 59}\Mach{18782}{14196 + 4586} \\
2b(ii), 29 & $25045$ & \Alt{\emph{(terminal stop)}} & $43391$ & $25045$ & $43389.56$ & \Mach{25045}{25045}\Mach{25045}{12529 + 12504 + 3\tms 4}\Mach{25043}{14196 + 4674 + 4586 + 800 + 2\tms 286 + \mbox{$3\tms 59 + 2\tms 19$}}\Mach{25043}{12550 + 4674 + 4620 + 918 + 2\tms 800 + 2\tms 286 + 67 + \mbox{$2\tms 19 + 4$}} \\
2b(iii), 27 & $12575$ & \Alt{$\le 5842$} & $30946$ & $17115$ & $29651.12$ & \Mach{17115}{12529 + 4586}\Mach{14632}{12550 + 918 + 800 + 286 + \mbox{$59 + 19$}}\Mach{14630}{2\tms 4674 + 4620 + 2\tms 286 + 59 + \mbox{$19 + 3\tms 4$}}\Mach{14629}{12575 + 2\tms 800 + 286 + 67 + 59 + \mbox{$2\tms 19 + 4$}} \\
2b(iii), 28 & $14210$ & \Alt{$\le 5842$} & $32581$ & $18805$ & $32578.99$ & \Mach{18805}{12575 + 4620 + 918 + 2\tms 286 + 59 + \mbox{$3\tms 19 + 4$}}\Mach{18804}{14210 + 4586 + 2\tms 4}\Mach{18804}{12550 + 4674 + 800 + 2\tms 286 + 67 + 2\tms 59 + \mbox{$19 + 4$}}\Mach{18803}{12529 + 4674 + 2\tms 800} \\
2b(iii), 29 & $25079$ & \Alt{\emph{(terminal stop)}} & $43450$ & $25079$ & $43448.47$ & \Mach{25079}{25079}\Mach{25079}{12550 + 12529}\Mach{25069}{14210 + 4674 + 4586 + 800 + 2\tms 286 + 67 + 2\tms 59 + \mbox{$2\tms 19 + 4$}}\Mach{25068}{12575 + 4674 + 4620 + 918 + 2\tms 800 + 2\tms 286 + 59 + \mbox{$2\tms 19 + 3\tms 4$}} \\
\addlinespace[5pt]
3a, 25 & $12413$ & \Alt{$5746 \to$ (i)}\Alt{$5788 \to$ (ii)--(iii)} & --- & $12413$ & --- & \Mach{12413}{12413}\Mach{9113}{4578 + 4535}\Mach{7019}{4620 + 918 + 800 + 2\tms 286 + 59 + \mbox{$2\tms 19 + 3\tms 4$}}\Mach{7019}{4620 + 2\tms 800 + 2\tms 286 + 67 + 2\tms 59 + \mbox{$2\tms 19 + 4$}} \\
3a(i), 26 & $12371$ & \Alt{$\le 5788$} & $30530$ & $12413$ & $21505.08$ & \Mach{12413}{12413}\Mach{12371}{12371}\Mach{11577}{4620 + 4535 + 918 + 800 + 2\tms 286 + 59 + \mbox{$3\tms 19 + 4\tms 4$}}\Mach{11574}{4620 + 4578 + 2\tms 800 + 2\tms 286 + 67 + \mbox{$2\tms 59 + 19$}} \\
3a(i), 27 & $12371$ & \Alt{$\le 5788$} & $30530$ & $16906$ & $29289.04$ & \Mach{16906}{12371 + 4535}\Mach{14468}{2\tms 4620 + 4578 + 2\tms 286 + 59 + 19}\Mach{14467}{12413 + 2\tms 800 + 286 + 67 + 59 + \mbox{$2\tms 19 + 4$}}\Mach{14465}{12371 + 918 + 800 + 286 + 59 + \mbox{$19 + 3\tms 4$}} \\
3a(i), 28 & $14051$ & \Alt{$\le 5788$} & $32210$ & $18591$ & $32208.24$ & \Mach{18591}{12371 + 4620 + 2\tms 800}\Mach{18590}{14051 + 4535 + 4}\Mach{18590}{12413 + 4578 + 918 + 2\tms 286 + 59 + \mbox{$2\tms 19 + 3\tms 4$}}\Mach{18586}{12371 + 4620 + 800 + 2\tms 286 + 67 + \mbox{$2\tms 59 + 2\tms 19$}} \\
3a(i), 29 & $24787$ & \Alt{\emph{(terminal stop)}} & $42946$ & $24787$ & $42942.59$ & \Mach{24787}{24787}\Mach{24787}{12413 + 4620 + 4578 + 918 + 2\tms 800 + 2\tms 286 + 59 + \mbox{$19 + 2\tms 4$}}\Mach{24786}{14051 + 4620 + 4535 + 800 + 2\tms 286 + 67 + 2\tms 59 + \mbox{$19 + 4$}}\Mach{24784}{2\tms 12371 + 2\tms 19 + 4} \\
3a(ii)--(iii), 26 & $12436$ & \Alt{$5746 \to$ (ii)}\Alt{$5788 \to$ (iii)} & $30637$ & $12436$ & $21544.92$ & \Mach{12436}{12436}\Mach{12413}{12413}\Mach{11577}{4620 + 4535 + 918 + 800 + 2\tms 286 + 59 + \mbox{$3\tms 19 + 4\tms 4$}}\Mach{11574}{4620 + 4578 + 2\tms 800 + 2\tms 286 + 67 + \mbox{$2\tms 59 + 19$}} \\
3a(ii), 27 & $12394$ & \Alt{$\le 5788$} & $30576$ & $16929$ & $29328.88$ & \Mach{16929}{12394 + 4535}\Mach{14490}{12436 + 2\tms 800 + 286 + 67 + 59 + \mbox{$2\tms 19 + 4$}}\Mach{14488}{12413 + 918 + 800 + 286 + \mbox{$59 + 3\tms 4$}}\Mach{14487}{2\tms 4620 + 4578 + 2\tms 286 + 59 + 2\tms 19} \\
3a(ii), 28 & $14074$ & \Alt{$\le 5788$} & $32256$ & $18618$ & $32255.02$ & \Mach{18618}{12413 + 4578 + 918 + 2\tms 286 + \mbox{$2\tms 59 + 19$}}\Mach{18618}{12394 + 4620 + 2\tms 800 + 4}\Mach{18617}{14074 + 4535 + 2\tms 4}\Mach{18615}{12436 + 4620 + 800 + 2\tms 286 + 67 + 59 + \mbox{$3\tms 19 + 4$}} \\
3a(ii), 29 & $24823$ & \Alt{\emph{(terminal stop)}} & $43005$ & $24823$ & $43004.96$ & \Mach{24823}{24823}\Mach{24823}{12436 + 2\tms 4620 + 918 + 2\tms 800 + \mbox{$2\tms 286 + 3\tms 19$}}\Mach{24823}{12413 + 12394 + 4\tms 4}\Mach{24822}{14074 + 4578 + 4535 + 800 + 2\tms 286 + 67 + \mbox{$3\tms 59 + 19$}} \\
3a(iii), 27 & $12455$ & \Alt{$\le 5788$} & $30656$ & $16948$ & $29361.80$ & \Mach{16948}{12413 + 4535}\Mach{14505}{12455 + 2\tms 800 + 286 + 67 + \mbox{$59 + 2\tms 19$}}\Mach{14503}{12436 + 918 + 800 + 286 + \mbox{$59 + 4$}}\Mach{14499}{2\tms 4620 + 4578 + 2\tms 286 + 59 + \mbox{$2\tms 19 + 3\tms 4$}} \\
3a(iii), 28 & $14091$ & \Alt{$\le 5788$} & $32292$ & $18639$ & $32291.40$ & \Mach{18639}{12455 + 4578 + 918 + 2\tms 286 + \mbox{$59 + 3\tms 19$}}\Mach{18637}{12413 + 4620 + 2\tms 800 + 4}\Mach{18636}{12436 + 4620 + 800 + 2\tms 286 + 67 + 2\tms 59 + \mbox{$19 + 4$}}\Mach{18634}{14091 + 4535 + 2\tms 4} \\
3a(iii), 29 & $24849$ & \Alt{\emph{(terminal stop)}} & $\mathbf{43050}$ & $24849$ & $\mathbf{43050.00}$ & \Mach{24849}{24849}\Mach{24849}{14091 + 4620 + 4535 + 800 + 2\tms 286 + 67 + 2\tms 59 + \mbox{$2\tms 19 + 2\tms 4$}}\Mach{24849}{12436 + 12413}\Mach{24848}{12455 + 4620 + 4578 + 918 + 2\tms 800 + 2\tms 286 + 59 + \mbox{$2\tms 19 + 2\tms 4$}} \\
\addlinespace[5pt]
3b, 25 & $12427$ & \Alt{$5754 \to$ (i)}\Alt{$5796 \to$ (ii)--(iii)} & --- & $12427$ & --- & \Mach{12427}{12427}\Mach{9132}{4586 + 4546}\Mach{7023}{4628 + 2\tms 800 + 2\tms 286 + 67 + \mbox{$2\tms 59 + 2\tms 19$}}\Mach{7023}{4620 + 918 + 800 + 2\tms 286 + 59 + \mbox{$2\tms 19 + 4\tms 4$}} \\
3b(i), 26 & $12385$ & \Alt{$\le 5796$} & $30566$ & $12427$ & $21529.33$ & \Mach{12427}{12427}\Mach{12385}{12385}\Mach{11590}{4620 + 4586 + 2\tms 800 + 2\tms 286 + 67 + 2\tms 59 + \mbox{$19 + 2\tms 4$}}\Mach{11588}{4628 + 4546 + 918 + 800 + 2\tms 286 + 59 + \mbox{$3\tms 19 + 2\tms 4$}} \\
3b(i), 27 & $12385$ & \Alt{$\le 5796$} & $30566$ & $16931$ & $29332.35$ & \Mach{16931}{12385 + 4546}\Mach{14484}{4628 + 4620 + 4586 + 2\tms 286 + \mbox{$59 + 19$}}\Mach{14481}{12427 + 2\tms 800 + 286 + 67 + 59 + \mbox{$2\tms 19 + 4$}}\Mach{14479}{12385 + 918 + 800 + 286 + 59 + \mbox{$19 + 3\tms 4$}} \\
3b(i), 28 & $14064$ & \Alt{$\le 5796$} & $32245$ & $18612$ & $32244.62$ & \Mach{18612}{12427 + 4586 + 918 + 2\tms 286 + 59 + \mbox{$2\tms 19 + 3\tms 4$}}\Mach{18610}{14064 + 4546}\Mach{18609}{12385 + 4620 + 2\tms 800 + 4}\Mach{18608}{12385 + 4628 + 800 + 2\tms 286 + 67 + \mbox{$2\tms 59 + 2\tms 19$}} \\
3b(i), 29 & $24814$ & \Alt{\emph{(terminal stop)}} & $42995$ & $24814$ & $42989.36$ & \Mach{24814}{24814}\Mach{24814}{14064 + 4620 + 4546 + 800 + 2\tms 286 + 67 + 2\tms 59 + \mbox{$19 + 2\tms 4$}}\Mach{24813}{12427 + 4628 + 4586 + 918 + 2\tms 800 + 2\tms 286 + 59 + \mbox{$19 + 4$}}\Mach{24812}{2\tms 12385 + 2\tms 19 + 4} \\
3b(ii)--(iii), 26 & $12450$ & \Alt{$5754 \to$ (ii)}\Alt{$5796 \to$ (iii)} & $30673$ & $12450$ & $21569.18$ & \Mach{12450}{12450}\Mach{12427}{12427}\Mach{11590}{4620 + 4586 + 2\tms 800 + 2\tms 286 + 67 + 2\tms 59 + \mbox{$19 + 2\tms 4$}}\Mach{11588}{4628 + 4546 + 918 + 800 + 2\tms 286 + 59 + \mbox{$3\tms 19 + 2\tms 4$}} \\
3b(ii), 27 & $12408$ & \Alt{$\le 5796$} & $30612$ & $16954$ & $29372.20$ & \Mach{16954}{12408 + 4546}\Mach{14504}{12450 + 2\tms 800 + 286 + 67 + 59 + \mbox{$2\tms 19 + 4$}}\Mach{14503}{4628 + 4620 + 4586 + 2\tms 286 + \mbox{$59 + 2\tms 19$}}\Mach{14502}{12427 + 918 + 800 + 286 + \mbox{$59 + 3\tms 4$}} \\
3b(ii), 28 & $14089$ & \Alt{$\le 5796$} & $32293$ & $18639$ & $32291.40$ & \Mach{18639}{14089 + 4546 + 4}\Mach{18639}{12450 + 4586 + 918 + 2\tms 286 + 67 + \mbox{$2\tms 19 + 2\tms 4$}}\Mach{18638}{12427 + 4620 + 800 + 2\tms 286 + 3\tms 59 + \mbox{$2\tms 19 + 4$}}\Mach{18636}{12408 + 4628 + 2\tms 800} \\
3b(ii), 29 & $24851$ & \Alt{\emph{(terminal stop)}} & $43055$ & $24851$ & $43053.46$ & \Mach{24851}{24851}\Mach{24851}{12450 + 4628 + 4586 + 918 + 2\tms 800 + 2\tms 286 + \mbox{$59 + 2\tms 19$}}\Mach{24851}{12427 + 12408 + 4\tms 4}\Mach{24850}{14089 + 4620 + 4546 + 800 + 2\tms 286 + 67 + \mbox{$2\tms 59 + 2\tms 19$}} \\
3b(iii), 27 & $12469$ & \Alt{$\le 5796$} & $30692$ & $16973$ & $29405.11$ & \Mach{16973}{12427 + 4546}\Mach{14519}{12469 + 2\tms 800 + 286 + 67 + \mbox{$59 + 2\tms 19$}}\Mach{14517}{12450 + 918 + 800 + 286 + \mbox{$59 + 4$}}\Mach{14515}{4628 + 4620 + 4586 + 2\tms 286 + 59 + \mbox{$2\tms 19 + 3\tms 4$}} \\
3b(iii), 28 & $14103$ & \Alt{$\le 5796$} & $32326$ & $18658$ & $32324.31$ & \Mach{18658}{12469 + 4586 + 918 + 2\tms 286 + 67 + \mbox{$2\tms 19 + 2\tms 4$}}\Mach{18657}{14103 + 4546 + 2\tms 4}\Mach{18657}{12450 + 4620 + 800 + 2\tms 286 + \mbox{$3\tms 59 + 2\tms 19$}}\Mach{18655}{12427 + 4628 + 2\tms 800} \\
3b(iii), 29 & $24877$ & \Alt{\emph{(terminal stop)}} & $43100$ & $24877$ & $43098.51$ & \Mach{24877}{24877}\Mach{24877}{12450 + 12427}\Mach{24876}{14103 + 4620 + 4546 + 800 + 2\tms 286 + 67 + 2\tms 59 + \mbox{$2\tms 19 + 3\tms 4$}}\Mach{24874}{12469 + 4628 + 4586 + 918 + 2\tms 800 + 2\tms 286 + 59 + \mbox{$2\tms 19 + 4$}} \\
\end{longtable}
}

%% file: tables/tab05_opt_phase3b.tex
{\scriptsize
\medmuskip=3mu\relax 
\relpenalty=10000\relax 
\setlength{\tabcolsep}{3.1pt}
\renewcommand{\arraystretch}{1.08}
\setlength{\LTcapwidth}{\textwidth}
\begin{longtable}{@{}>{\raggedright\arraybackslash}p{0.115\textwidth} r >{\raggedright\arraybackslash}p{0.172\textwidth} r r r >{\raggedright\arraybackslash}p{0.402\textwidth}@{}}
\caption{\Cref{thm:opt}, Phase~3 endgame (Steps 25--29) for Cases 4a, 4b, 4c, and 4d across Subcases (i), (ii), and (iii), starting from the reduced Phase-$2$ loads $(h, h, h, l)$ of \Cref{tab:opt_strategy}(b).}\label{tab:phase3b}\\
\toprule
Case, Step & Job & Valid / Transition & $\ALG \ge$ & $\OPT \le$ & $c \cdot \OPT$ & Offline witness schedule ($\mathbf{load}:\text{jobs}$) \\
\midrule
\endfirsthead
\multicolumn{7}{@{}l}{\small\emph{\Cref{tab:phase3b} (continued)}}\\[2pt]
\toprule
Case, Step & Job & Valid / Transition & $\ALG \ge$ & $\OPT \le$ & $c \cdot \OPT$ & Offline witness schedule ($\mathbf{load}:\text{jobs}$) \\
\midrule
\endhead
\midrule
\multicolumn{7}{r@{}}{\emph{Continued on next page}}\\
\endfoot
\bottomrule
\endlastfoot
4a, 25 & $12321$ & \Alt{$5703 \to$ (i)}\Alt{$5746 \to$ (ii)--(iii)} & --- & $12321$ & --- & \Mach{12321}{12321}\Mach{8987}{4535 + 4452}\Mach{7000}{4620 + 2\tms 800 + 2\tms 286 + 67 + 2\tms 59 + \mbox{$19 + 4$}}\Mach{6996}{4578 + 918 + 800 + 2\tms 286 + 59 + \mbox{$3\tms 19 + 3\tms 4$}} \\
4a(i), 26 & $12278$ & \Alt{$\le 5746$} & $30302$ & $12321$ & $21345.69$ & \Mach{12321}{12321}\Mach{12278}{12278}\Mach{11493}{4578 + 4535 + 2\tms 800 + 2\tms 286 + 67 + 2\tms 59 + \mbox{$19 + 4$}}\Mach{11490}{4620 + 4452 + 918 + 800 + 2\tms 286 + 59 + \mbox{$3\tms 19 + 3\tms 4$}} \\
4a(i), 27 & $12278$ & \Alt{$\le 5746$} & $30302$ & $16730$ & $28984.12$ & \Mach{16730}{12278 + 4452}\Mach{14379}{12278 + 918 + 800 + 286 + \mbox{$59 + 2\tms 19$}}\Mach{14376}{4620 + 4578 + 4535 + 2\tms 286 + \mbox{$59 + 3\tms 4$}}\Mach{14375}{12321 + 2\tms 800 + 286 + 67 + 59 + \mbox{$2\tms 19 + 4$}} \\
4a(i), 28 & $13951$ & \Alt{$\le 5746$} & $31975$ & $18456$ & $31974.36$ & \Mach{18456}{12278 + 4578 + 2\tms 800}\Mach{18455}{12278 + 4620 + 800 + 2\tms 286 + \mbox{$67 + 2\tms 59$}}\Mach{18451}{12321 + 4535 + 918 + 2\tms 286 + 59 + \mbox{$2\tms 19 + 2\tms 4$}}\Mach{18449}{13951 + 4452 + 2\tms 19 + 2\tms 4} \\
4a(i), 29 & $24604$ & \Alt{\emph{(terminal stop)}} & $42628$ & $24605$ & $42627.28$ & \Mach{24605}{13951 + 4578 + 4452 + 800 + 2\tms 286 + 67 + \mbox{$3\tms 59 + 2\tms 4$}}\Mach{24604}{24604}\Mach{24604}{12321 + 4620 + 4535 + 918 + 2\tms 800 + \mbox{$2\tms 286 + 2\tms 19$}}\Mach{24602}{2\tms 12278 + 2\tms 19 + 2\tms 4} \\
4a(ii)--(iii), 26 & $12345$ & \Alt{$5703 \to$ (ii)}\Alt{$5746 \to$ (iii)} & $30412$ & $12345$ & $21387.27$ & \Mach{12345}{12345}\Mach{12321}{12321}\Mach{11493}{4578 + 4535 + 2\tms 800 + 2\tms 286 + 67 + 2\tms 59 + \mbox{$19 + 4$}}\Mach{11490}{4620 + 4452 + 918 + 800 + 2\tms 286 + 59 + \mbox{$3\tms 19 + 3\tms 4$}} \\
4a(ii), 27 & $12302$ & \Alt{$\le 5746$} & $30350$ & $16754$ & $29025.70$ & \Mach{16754}{12302 + 4452}\Mach{14402}{4620 + 4578 + 4535 + 2\tms 286 + \mbox{$59 + 2\tms 19$}}\Mach{14399}{12345 + 2\tms 800 + 286 + 67 + 59 + \mbox{$2\tms 19 + 4$}}\Mach{14396}{12321 + 918 + 800 + 286 + \mbox{$59 + 3\tms 4$}} \\
4a(ii), 28 & $13968$ & \Alt{$\le 5746$} & $32016$ & $18480$ & $32015.94$ & \Mach{18480}{12345 + 4620 + 800 + 2\tms 286 + \mbox{$67 + 4\tms 19$}}\Mach{18480}{12321 + 4535 + 918 + 2\tms 286 + \mbox{$2\tms 59 + 4\tms 4$}}\Mach{18480}{12302 + 4578 + 2\tms 800}\Mach{18479}{13968 + 4452 + 59} \\
4a(ii), 29 & $24640$ & \Alt{\emph{(terminal stop)}} & $42688$ & $24640$ & $42687.92$ & \Mach{24640}{24640}\Mach{24640}{13968 + 4620 + 4452 + 2\tms 800}\Mach{24640}{12345 + 4578 + 4535 + 918 + 800 + 4\tms 286 + 67 + \mbox{$3\tms 59 + 4\tms 19$}}\Mach{24639}{12321 + 12302 + 4\tms 4} \\
4a(iii), 27 & $12364$ & \Alt{$\le 5746$} & $30431$ & $16773$ & $29058.62$ & \Mach{16773}{12321 + 4452}\Mach{14414}{12364 + 2\tms 800 + 286 + 67 + \mbox{$59 + 2\tms 19$}}\Mach{14414}{4620 + 4578 + 4535 + 2\tms 286 + 59 + \mbox{$2\tms 19 + 3\tms 4$}}\Mach{14412}{12345 + 918 + 800 + 286 + \mbox{$59 + 4$}} \\
4a(iii), 28 & $13982$ & \Alt{$\le 5746$} & $32049$ & $18499$ & $32048.85$ & \Mach{18499}{13982 + 4452 + 3\tms 19 + 2\tms 4}\Mach{18499}{12364 + 4535 + 2\tms 800}\Mach{18499}{12345 + 4578 + 918 + 2\tms 286 + 59 + \mbox{$19 + 2\tms 4$}}\Mach{18498}{12321 + 4620 + 800 + 2\tms 286 + \mbox{$67 + 2\tms 59$}} \\
4a(iii), 29 & $24666$ & \Alt{\emph{(terminal stop)}} & $42733$ & $24666$ & $42732.96$ & \Mach{24666}{24666}\Mach{24666}{13982 + 4578 + 4452 + 800 + 2\tms 286 + 67 + \mbox{$3\tms 59 + 2\tms 19$}}\Mach{24666}{12345 + 12321}\Mach{24663}{12364 + 4620 + 4535 + 918 + 2\tms 800 + 2\tms 286 + \mbox{$2\tms 19 + 4\tms 4$}} \\
\addlinespace[5pt]
4b, 25 & $12323$ & \Alt{$5711 \to$ (i)}\Alt{$5746 \to$ (ii)--(iii)} & --- & $12323$ & --- & \Mach{12323}{12323}\Mach{8995}{4535 + 4460}\Mach{7000}{4620 + 2\tms 800 + 2\tms 286 + 67 + 2\tms 59 + \mbox{$19 + 4$}}\Mach{6996}{4578 + 918 + 800 + 2\tms 286 + 59 + \mbox{$3\tms 19 + 3\tms 4$}} \\
4b(i), 26 & $12288$ & \Alt{$\le 5746$} & $30322$ & $12323$ & $21349.15$ & \Mach{12323}{12323}\Mach{12288}{12288}\Mach{11497}{4578 + 4535 + 2\tms 800 + 2\tms 286 + 67 + 2\tms 59 + \mbox{$19 + 2\tms 4$}}\Mach{11494}{4620 + 4460 + 918 + 800 + 2\tms 286 + 59 + \mbox{$3\tms 19 + 2\tms 4$}} \\
4b(i), 27 & $12288$ & \Alt{$\le 5746$} & $30322$ & $16748$ & $29015.31$ & \Mach{16748}{12288 + 4460}\Mach{14383}{4620 + 4578 + 4535 + 2\tms 286 + \mbox{$59 + 19$}}\Mach{14381}{12323 + 2\tms 800 + 286 + 67 + 59 + \mbox{$2\tms 19 + 2\tms 4$}}\Mach{14378}{12288 + 918 + 800 + 286 + 59 + \mbox{$19 + 2\tms 4$}} \\
4b(i), 28 & $13955$ & \Alt{$\le 5746$} & $31989$ & $18462$ & $31984.75$ & \Mach{18462}{12323 + 4535 + 2\tms 800 + 4}\Mach{18461}{13955 + 4460 + 2\tms 19 + 2\tms 4}\Mach{18461}{12288 + 4620 + 800 + 2\tms 286 + \mbox{$3\tms 59 + 4$}}\Mach{18461}{12288 + 4578 + 918 + 2\tms 286 + \mbox{$67 + 2\tms 19$}} \\
4b(i), 29 & $24615$ & \Alt{\emph{(terminal stop)}} & $42649$ & $24617$ & $42648.07$ & \Mach{24617}{13955 + 4578 + 4460 + 800 + 2\tms 286 + 67 + \mbox{$3\tms 59 + 2\tms 4$}}\Mach{24615}{24615}\Mach{24614}{12323 + 4620 + 4535 + 918 + 2\tms 800 + 2\tms 286 + \mbox{$2\tms 19 + 2\tms 4$}}\Mach{24614}{2\tms 12288 + 2\tms 19} \\
4b(ii)--(iii), 26 & $12344$ & \Alt{$5711 \to$ (ii)}\Alt{$5746 \to$ (iii)} & $30413$ & $12344$ & $21385.54$ & \Mach{12344}{12344}\Mach{12323}{12323}\Mach{11497}{4578 + 4535 + 2\tms 800 + 2\tms 286 + 67 + 2\tms 59 + \mbox{$19 + 2\tms 4$}}\Mach{11494}{4620 + 4460 + 918 + 800 + 2\tms 286 + 59 + \mbox{$3\tms 19 + 2\tms 4$}} \\
4b(ii), 27 & $12309$ & \Alt{$\le 5746$} & $30364$ & $16769$ & $29051.69$ & \Mach{16769}{12309 + 4460}\Mach{14402}{4620 + 4578 + 4535 + 2\tms 286 + \mbox{$59 + 2\tms 19$}}\Mach{14398}{12344 + 2\tms 800 + 286 + 67 + 59 + \mbox{$2\tms 19 + 4$}}\Mach{14398}{12323 + 918 + 800 + 286 + \mbox{$59 + 3\tms 4$}} \\
4b(ii), 28 & $13974$ & \Alt{$\le 5746$} & $32029$ & $18487$ & $32028.06$ & \Mach{18487}{12344 + 4535 + 918 + 2\tms 286 + 2\tms 59}\Mach{18487}{12309 + 4578 + 2\tms 800}\Mach{18484}{13974 + 4460 + 2\tms 19 + 3\tms 4}\Mach{18483}{12323 + 4620 + 800 + 2\tms 286 + 67 + 59 + \mbox{$2\tms 19 + 4$}} \\
4b(ii), 29 & $24648$ & \Alt{\emph{(terminal stop)}} & $42703$ & $24648$ & $42701.77$ & \Mach{24648}{24648}\Mach{24648}{12323 + 12309 + 4\tms 4}\Mach{24647}{13974 + 4578 + 4460 + 800 + 2\tms 286 + 67 + \mbox{$3\tms 59 + 19$}}\Mach{24646}{12344 + 4620 + 4535 + 918 + 2\tms 800 + \mbox{$2\tms 286 + 3\tms 19$}} \\
4b(iii), 27 & $12358$ & \Alt{$\le 5746$} & $30427$ & $16783$ & $29075.94$ & \Mach{16783}{12323 + 4460}\Mach{14412}{12358 + 2\tms 800 + 286 + 67 + 59 + \mbox{$2\tms 19 + 4$}}\Mach{14411}{12344 + 918 + 800 + 286 + \mbox{$59 + 4$}}\Mach{14410}{4620 + 4578 + 4535 + 2\tms 286 + 59 + \mbox{$2\tms 19 + 2\tms 4$}} \\
4b(iii), 28 & $13985$ & \Alt{$\le 5746$} & $32054$ & $18501$ & $32052.32$ & \Mach{18501}{12358 + 4535 + 918 + 2\tms 286 + 2\tms 59}\Mach{18501}{12323 + 4578 + 2\tms 800}\Mach{18500}{12344 + 4620 + 800 + 2\tms 286 + 67 + \mbox{$59 + 2\tms 19$}}\Mach{18499}{13985 + 4460 + 2\tms 19 + 4\tms 4} \\
4b(iii), 29 & $24667$ & \Alt{\emph{(terminal stop)}} & $42736$ & $24667$ & $42734.69$ & \Mach{24667}{24667}\Mach{24667}{13985 + 4578 + 4535 + 800 + 2\tms 286 + 67 + \mbox{$2\tms 59 + 3\tms 4$}}\Mach{24667}{12358 + 4620 + 4460 + 918 + 2\tms 800 + 2\tms 286 + 59 + \mbox{$4\tms 19 + 4$}}\Mach{24667}{12344 + 12323} \\
\addlinespace[5pt]
4c, 25 & $12330$ & \Alt{$5703 \to$ (i)}\Alt{$5754 \to$ (ii)--(iii)} & --- & $12330$ & --- & \Mach{12330}{12330}\Mach{9003}{4535 + 4468}\Mach{7000}{4620 + 2\tms 800 + 2\tms 286 + 67 + 2\tms 59 + \mbox{$19 + 4$}}\Mach{6996}{4578 + 918 + 800 + 2\tms 286 + 59 + \mbox{$3\tms 19 + 3\tms 4$}} \\
4c(i), 26 & $12279$ & \Alt{$\le 5754$} & $30312$ & $12330$ & $21361.28$ & \Mach{12330}{12330}\Mach{12279}{12279}\Mach{11501}{4578 + 4535 + 2\tms 800 + 2\tms 286 + 67 + 2\tms 59 + \mbox{$19 + 3\tms 4$}}\Mach{11498}{4620 + 4468 + 918 + 800 + 2\tms 286 + 59 + \mbox{$3\tms 19 + 4$}} \\
4c(i), 27 & $12279$ & \Alt{$\le 5754$} & $30312$ & $16747$ & $29013.58$ & \Mach{16747}{12279 + 4468}\Mach{14380}{12330 + 2\tms 800 + 286 + 67 + \mbox{$59 + 2\tms 19$}}\Mach{14380}{12279 + 918 + 800 + 286 + \mbox{$59 + 2\tms 19$}}\Mach{14380}{4620 + 4578 + 4535 + 2\tms 286 + \mbox{$59 + 4\tms 4$}} \\
4c(i), 28 & $13954$ & \Alt{$\le 5754$} & $31987$ & $18461$ & $31983.02$ & \Mach{18461}{12279 + 4578 + 2\tms 800 + 4}\Mach{18460}{13954 + 4468 + 2\tms 19}\Mach{18460}{12330 + 4535 + 918 + 2\tms 286 + 59 + \mbox{$2\tms 19 + 2\tms 4$}}\Mach{18460}{12279 + 4620 + 800 + 2\tms 286 + 67 + \mbox{$2\tms 59 + 4$}} \\
4c(i), 29 & $24614$ & \Alt{\emph{(terminal stop)}} & $42647$ & $24616$ & $42646.34$ & \Mach{24616}{13954 + 4578 + 4468 + 800 + 2\tms 286 + \mbox{$67 + 3\tms 59$}}\Mach{24614}{24614}\Mach{24613}{12330 + 4620 + 4535 + 918 + 2\tms 800 + \mbox{$2\tms 286 + 2\tms 19$}}\Mach{24612}{2\tms 12279 + 2\tms 19 + 4\tms 4} \\
4c(ii)--(iii), 26 & $12359$ & \Alt{$5703 \to$ (ii)}\Alt{$5754 \to$ (iii)} & $30443$ & $12359$ & $21411.52$ & \Mach{12359}{12359}\Mach{12330}{12330}\Mach{11501}{4578 + 4535 + 2\tms 800 + 2\tms 286 + 67 + 2\tms 59 + \mbox{$19 + 3\tms 4$}}\Mach{11498}{4620 + 4468 + 918 + 800 + 2\tms 286 + 59 + \mbox{$3\tms 19 + 4$}} \\
4c(ii), 27 & $12308$ & \Alt{$\le 5754$} & $30370$ & $16776$ & $29063.82$ & \Mach{16776}{12308 + 4468}\Mach{14409}{12359 + 2\tms 800 + 286 + 67 + \mbox{$59 + 2\tms 19$}}\Mach{14406}{4620 + 4578 + 4535 + 2\tms 286 + 59 + \mbox{$2\tms 19 + 4$}}\Mach{14405}{12330 + 918 + 800 + 286 + \mbox{$59 + 3\tms 4$}} \\
4c(ii), 28 & $13979$ & \Alt{$\le 5754$} & $32041$ & $18494$ & $32040.19$ & \Mach{18494}{12359 + 4535 + 2\tms 800}\Mach{18494}{12330 + 4620 + 800 + 2\tms 286 + 67 + 59 + \mbox{$2\tms 19 + 2\tms 4$}}\Mach{18494}{12308 + 4578 + 918 + 2\tms 286 + 2\tms 59}\Mach{18493}{13979 + 4468 + 2\tms 19 + 2\tms 4} \\
4c(ii), 29 & $24659$ & \Alt{\emph{(terminal stop)}} & $42721$ & $24659$ & $42720.83$ & \Mach{24659}{24659}\Mach{24659}{13979 + 4578 + 4535 + 800 + 2\tms 286 + 67 + 59 + \mbox{$3\tms 19 + 3\tms 4$}}\Mach{24659}{12359 + 4620 + 4468 + 918 + 2\tms 800 + 2\tms 286 + \mbox{$2\tms 59 + 4$}}\Mach{24657}{12330 + 12308 + 19} \\
4c(iii), 27 & $12381$ & \Alt{$\le 5754$} & $30465$ & $16798$ & $29101.93$ & \Mach{16798}{12330 + 4468}\Mach{14425}{4620 + 4578 + 4535 + 2\tms 286 + 59 + \mbox{$3\tms 19 + 4$}}\Mach{14424}{12381 + 2\tms 800 + 286 + 67 + 59 + \mbox{$19 + 3\tms 4$}}\Mach{14422}{12359 + 918 + 800 + 286 + 59} \\
4c(iii), 28 & $13998$ & \Alt{$\le 5754$} & $32082$ & $18518$ & $32081.77$ & \Mach{18518}{12330 + 4620 + 800 + 2\tms 286 + \mbox{$3\tms 59 + 19$}}\Mach{18517}{12359 + 4578 + 918 + 2\tms 286 + 67 + \mbox{$19 + 4$}}\Mach{18516}{13998 + 4468 + 2\tms 19 + 3\tms 4}\Mach{18516}{12381 + 4535 + 2\tms 800} \\
4c(iii), 29 & $24689$ & \Alt{\emph{(terminal stop)}} & $42773$ & $24689$ & $42772.81$ & \Mach{24689}{24689}\Mach{24689}{13998 + 4578 + 4468 + 800 + 2\tms 286 + 67 + 2\tms 59 + \mbox{$4\tms 19 + 3\tms 4$}}\Mach{24689}{12381 + 4620 + 4535 + 918 + 2\tms 800 + 2\tms 286 + \mbox{$59 + 4$}}\Mach{24689}{12359 + 12330} \\
\addlinespace[5pt]
4d, 25 & $12339$ & \Alt{$5714 \to$ (i)}\Alt{$5754 \to$ (ii)--(iii)} & --- & $12339$ & --- & \Mach{12339}{12339}\Mach{9014}{4546 + 4468}\Mach{7004}{4620 + 2\tms 800 + 2\tms 286 + 67 + 2\tms 59 + \mbox{$19 + 2\tms 4$}}\Mach{7000}{4586 + 918 + 800 + 2\tms 286 + 59 + \mbox{$3\tms 19 + 2\tms 4$}} \\
4d(i), 26 & $12299$ & \Alt{$\le 5754$} & $30352$ & $12339$ & $21376.87$ & \Mach{12339}{12339}\Mach{12299}{12299}\Mach{11510}{4620 + 4468 + 918 + 800 + 2\tms 286 + 59 + \mbox{$3\tms 19 + 4\tms 4$}}\Mach{11508}{4586 + 4546 + 2\tms 800 + 2\tms 286 + 67 + \mbox{$2\tms 59 + 19$}} \\
4d(i), 27 & $12299$ & \Alt{$\le 5754$} & $30352$ & $16767$ & $29048.23$ & \Mach{16767}{12299 + 4468}\Mach{14400}{12299 + 918 + 800 + 286 + \mbox{$59 + 2\tms 19$}}\Mach{14395}{4620 + 4586 + 4546 + 2\tms 286 + \mbox{$59 + 3\tms 4$}}\Mach{14393}{12339 + 2\tms 800 + 286 + 67 + 59 + \mbox{$2\tms 19 + 4$}} \\
4d(i), 28 & $13971$ & \Alt{$\le 5754$} & $32024$ & $18483$ & $32021.13$ & \Mach{18483}{12299 + 4546 + 2\tms 800 + 2\tms 19}\Mach{18482}{12339 + 4586 + 800 + 2\tms 286 + \mbox{$67 + 2\tms 59$}}\Mach{18481}{13971 + 4468 + 2\tms 19 + 4}\Mach{18480}{12299 + 4620 + 918 + 2\tms 286 + \mbox{$59 + 3\tms 4$}} \\
4d(i), 29 & $24642$ & \Alt{\emph{(terminal stop)}} & $42695$ & $24644$ & $42694.84$ & \Mach{24644}{2\tms 12299 + 2\tms 19 + 2\tms 4}\Mach{24642}{24642}\Mach{24641}{13971 + 4586 + 4468 + 800 + 2\tms 286 + \mbox{$67 + 3\tms 59$}}\Mach{24641}{12339 + 4620 + 4546 + 918 + 2\tms 800 + 2\tms 286 + \mbox{$2\tms 19 + 2\tms 4$}} \\
4d(ii)--(iii), 26 & $12362$ & \Alt{$5714 \to$ (ii)}\Alt{$5754 \to$ (iii)} & $30455$ & $12362$ & $21416.72$ & \Mach{12362}{12362}\Mach{12339}{12339}\Mach{11510}{4620 + 4468 + 918 + 800 + 2\tms 286 + 59 + \mbox{$3\tms 19 + 4\tms 4$}}\Mach{11508}{4586 + 4546 + 2\tms 800 + 2\tms 286 + 67 + \mbox{$2\tms 59 + 19$}} \\
4d(ii), 27 & $12322$ & \Alt{$\le 5754$} & $30398$ & $16790$ & $29088.07$ & \Mach{16790}{12322 + 4468}\Mach{14421}{12339 + 918 + 800 + 286 + \mbox{$59 + 19$}}\Mach{14416}{12362 + 2\tms 800 + 286 + 67 + 59 + \mbox{$2\tms 19 + 4$}}\Mach{14414}{4620 + 4586 + 4546 + 2\tms 286 + 59 + \mbox{$19 + 3\tms 4$}} \\
4d(ii), 28 & $13990$ & \Alt{$\le 5754$} & $32066$ & $18508$ & $32064.45$ & \Mach{18508}{13990 + 4468 + 2\tms 19 + 3\tms 4}\Mach{18508}{12339 + 4620 + 800 + 2\tms 286 + 3\tms 59}\Mach{18508}{12322 + 4586 + 2\tms 800}\Mach{18507}{12362 + 4546 + 918 + 2\tms 286 + 67 + \mbox{$2\tms 19 + 4$}} \\
4d(ii), 29 & $24677$ & \Alt{\emph{(terminal stop)}} & $42753$ & $24677$ & $42752.02$ & \Mach{24677}{24677}\Mach{24677}{13990 + 4586 + 4468 + 800 + 2\tms 286 + 67 + \mbox{$2\tms 59 + 4\tms 19$}}\Mach{24677}{12362 + 4620 + 4546 + 918 + 2\tms 800 + \mbox{$2\tms 286 + 59$}}\Mach{24677}{12339 + 12322 + 4\tms 4} \\
4d(iii), 27 & $12379$ & \Alt{$\le 5754$} & $30472$ & $16807$ & $29117.52$ & \Mach{16807}{12339 + 4468}\Mach{14433}{12379 + 2\tms 800 + 286 + 67 + 59 + \mbox{$2\tms 19 + 4$}}\Mach{14429}{12362 + 918 + 800 + 286 + \mbox{$59 + 4$}}\Mach{14429}{4620 + 4586 + 4546 + 2\tms 286 + 59 + \mbox{$2\tms 19 + 2\tms 4$}} \\
4d(iii), 28 & $14005$ & \Alt{$\le 5754$} & $32098$ & $18527$ & $32097.36$ & \Mach{18527}{14005 + 4468 + 2\tms 19 + 4\tms 4}\Mach{18527}{12339 + 4620 + 800 + 2\tms 286 + \mbox{$3\tms 59 + 19$}}\Mach{18525}{12379 + 4546 + 2\tms 800}\Mach{18524}{12362 + 4586 + 918 + 2\tms 286 + \mbox{$67 + 19$}} \\
4d(iii), 29 & $24701$ & \Alt{\emph{(terminal stop)}} & $42794$ & $24701$ & $42793.60$ & \Mach{24701}{24701}\Mach{24701}{14005 + 4586 + 4468 + 2\tms 800 + \mbox{$2\tms 19 + 4$}}\Mach{24701}{12379 + 4620 + 4546 + 918 + 800 + 4\tms 286 + 67 + 3\tms 59 + \mbox{$2\tms 19 + 3\tms 4$}}\Mach{24701}{12362 + 12339} \\
\end{longtable}
}